\documentclass[iop,apj,tighten]{emulateapj}

\shorttitle{HR 8799 Astrometry and Orbits}
\shortauthors{Konopacky et al.}
\begin{document}

\title{Astrometric Monitoring of the HR 8799 Planets: Orbit
  Constraints from Self-Consistent Measurements}
\author{Q.M. Konopacky\altaffilmark{1},
  C. Marois\altaffilmark{2}, B.A. Macintosh\altaffilmark{3}, R. Galicher\altaffilmark{4,5},
  T.S. Barman\altaffilmark{6}, S.A. Metchev\altaffilmark{7,8},
  B. Zuckerman\altaffilmark{9}}
\altaffiltext{1}{Center for Astrophysics and Space Sciences,
  University of California, San Diego, La
  Jolla, CA 92093, USA; qkonopacky@ucsd.edu} 
\altaffiltext{2}{National Research Council Canada, Dominion
  Astrophysical Observatory, Victoria,
  BC V9E 2E7, Canada; christian.marois@nrc-cnrc.gc.ca} 
\altaffiltext{3}{Kavli Institute for Particle Astrophysics and
  Cosmology, Stanford University, Stanford, CA 94305, USA; bmacintosh@stanford.edu}
\altaffiltext{4}{LESIA, Observatoire de Paris, CNRS,
  Universit\'e Paris Diderot, Universit\'e Pierre et Marie
  Curie, 5 place Jules Janssen, 92190 Meudon, France;
  raphael.galicher@obspm.fr} 
\altaffiltext{5}{Groupement d'int\'er\^et Scientifique PHASE
  (Partenariat Haute r\'esolution Angulaire Sol Espace)
  between ONERA, Observatoire de Paris, CNRS and Universit\'e
  Paris Diderot} 
\altaffiltext{6}{Lunar and Planetary Lab, University of
  Arizona, Tucson, AZ 85721, USA; barman@lpl.arizona.edu} 
\altaffiltext{7}{Department of Physics and Astronomy, Centre for
  Planetary Science and Exploration, The University of Western
  Ontario, London, ON N6A 3K7, Canada; smetchev@uwo.ca}
\altaffiltext{8}{Department of Physics and Astronomy, Stony
  Brook University, Stony Brook, NY 11794-3800, USA}
\altaffiltext{9}{Department of Physics and Astronomy,
  University of California, Los Angeles, Los Angeles, CA 90095, USA;
  ben@astro.ucla.edu}

\begin{abstract}
We present new astrometric measurements from our ongoing
monitoring campaign of the HR 8799 directly imaged planetary
system.  These new data points were obtained with NIRC2 on the
W.M. Keck II 10 meter telescope between 2009 and
2014.  In addition, we present updated astrometry from
previously published observations in 2007 and 2008.  All data
were reduced using the SOSIE algorithm, which accounts for
systematic biases present in previously published
observations.  This allows us to construct a self-consistent
data set derived entirely from NIRC2 data alone.  From this
dataset, we detect acceleration for
two of the planets (HR 8799b and e) at $>$3$\sigma$.  We also assess possible
orbital parameters for each of the four planets independently.
We find no
statistically significant difference in the allowed
inclinations of the planets.
Fitting the astrometry while forcing coplanarity also returns $\chi^2$ consistent to within
1$\sigma$ of the best fit values, suggesting that if
inclination offsets of $\lesssim$20$^{o}$ are present, they are not detectable
with current data.  Our orbital fits also favor low
eccentricities, consistent with predictions from dynamical
modeling.  We also find period distributions consistent to within
1$\sigma$ with a 1:2:4:8 resonance between all planets.  This analysis
demonstrates the importance of minimizing astrometric systematics when fitting for
solutions to highly undersampled orbits. 

\end{abstract}

\section{Introduction}

Direct imaging offers a powerful tool for the discovery
and characterization of Jovian exoplanets.  The currently
known directly imaged planets are generally massive ($\sim$2-10
M$_{J}$), residing at wide separations from their host stars
($\sim$10-200 AU, e.g.,
\citealt{marois08,kalas08,marois10a,lagrange09,rameau13,kuzuhara13,macintosh15}).
Formation theories do not naturally predict 
the existence of all directly imaged planets, leading to speculation that
dynamical interactions, such as scattering or migration, shortly after the formation of these
objects plays a key role in generating their currently
observed configuration (e.g., \citealt{crida09,veras09}).  In order to assess the
dynamical history of these objects, empirical constraints on their
current orbital parameters are required.

Given the wide separations of these planets, their predicted orbital periods are tens
to hundered of years.  This means that while there is some hope
of obtaining full orbital phase coverage for the
shortest period systems (for example, $\beta$ Pictoris b,
\citealt{chauvin12,mmb15}), generally we must be content with
fractional orbit coverage.  In 
order to make the most of partial orbital information, precise
relative astrometry between the host star and the planet is
essential.  Such data has been shown to yield
useful dynamical constraints in other astronomical
contexts (e.g., \citealt{duchene06,jlu09}), and can likely do the same for imaged planets.  

The HR 8799 planetary system offers one of the most
interesting laboratories for measuring dynamics in a directly
imaged system.  With four imaged planets (HR 8799b, c, d, and
e) ranging in projected separation from $\sim$15 to 70 AU
(\citealt{marois08, marois10a}), the
system presents the opportunity to empirically measure orbits
and assess the fidelity of those orbit predictions using
multiplanet dynamical simulations.  Since their original discovery with the
W.M. Keck telescope and the Gemini North telescope, the planets
in the system
system have now been observed by 13 independent high-contrast
imaging systems and telescopes, offering a complex,
multiwavelength dataset spanning 16 years (e.g,
\citealt{lafreniere09,fukagawa09,metchev09,hinz10,serabyn10,bergfors11,soummer11,skemer12,ingraham14,currie14,pueyo15,rajan15,zurlo15}). 

In exploring the dynamical stability of HR 8799,
\citet{fabrycky10} were the first to point out that a multiple
mean-motion 
Laplace resonance was essential to the long term stability of
the system given the high estimated masses of the planets
($\sim$4-10 M$_{Jup}$).
Similarly, \citet{reidemeister09} and \citet{moro10} found that a 1:2:4 resonance
was necessary for stability, in addition to a non-face on
orbital inclination for a three planet system.
With the addition of HR 8799e, \citet{marois10a} 
found that the masses of the planets are likely $\lesssim$7
M$_{Jup}$ based on the stable
solutions of \citet{fabrycky10} for a three-planet HR 8799 and
a younger system age \citep{zuckerman11}.  In a large scale
simulation, \citet{sudol12} similarly determined that the
masses must be $\lesssim$10 M$_{Jup}$.   More recently,
\citet{goz14} demonstrated that the four HR 8799 planets could have migrated
into their current configuration shortly after formation,
finding final orbital configurations that are consistent with
published astrometry given an inclination of 25$^{\circ}$.
They also predict the location of a putative additional interior
planet given stability requirements (either $\sim$7.5 AU or
$\sim$9.5 AU).

Several authors have used existing and new relative
astrometric measurements, which typically have a precision of
5-10 milliarcseconds (mas), to empirically constrain the possible
orbits of the four HR 8799 planets. In many cases, they have
used the results of the dynamical simulations described
previously as a starting point for orbit fitting.  For
example, in their recovery of HR 8799b, c, and d in archival
Hubble Space Telescope data from 1998, \citet{soummer11} tested
the 1:2:4 resonance hypothesis, assuming coplanarity, to place
constraints on the system inclination and 
eccentricity.  Similarly, \citet{currie12} included archival
data from Keck to assess the eccentricities and inclinations
for the system assuming a 1:2:4 period ratio, finding that
face-on orbits did not provide solutions consistent with these
periods.  They also found that HR 8799d appeared to be
non-coplanar with HR 8799b and c.  More recently,
\citet{esposito13} and \citet{maire15} have added astrometric
data points from the Large Binocular Telescope to constrain
possible orbits.  They find consistency with the 
1:2:4:8 mean motion resonance, but also inclination offsets for HR 8799d.
A similar analysis was recently performed by \citet{zurlo15}
using new data from SPHERE on the Very Large Telescope (VLT).
In looking at possible orbits consistent with all available
astrometry, \citet{zurlo15} conclude that HR 8799d and e may be in a 1:2
or 2:3 resonance rather than a possible 2:5 resonance.
With the addition of an astrometric data point in 2012 from
Project 1640, \citet{pueyo15} make no assumptions about the
orbital properties, instead fitting for all orbital parameters
with generously large priors.  They also find that HR 8799d has an inclination
differing from the other planets.  \citet{pueyo15} also assert
that the masses must be below 13 M$_{Jup}$, consistent with
previous dynamical analyses.  All of these works combine
astrometry from multiple telescopes and instruments, with a large fraction
of astrometric data coming from data taken with Keck.  

In this paper, we present new and updated astrometric
measurements for the four HR 8799 planets obtained with Keck
II.  By removing systematic biases in our astrometry, we are
able to construct distributions of potential orbits from a
fully \textit{self-consistent} data set.  In Section
\ref{data}, we describe our data and our improved reduction
methods.  In Section \ref{orbits}, we describe our method for orbit
fitting and summarize the orbit parameters allowed with our
new astrometry and our detection of acceleration in two of the
planets.  In Section \ref{discussion} we discuss our findings in the
context of other analyses and describe future measurements
that could further elucidate the dynamical history of this exoplanetary system.

\section{Astrometric Data and Analysis}\label{data}

Imaging data from which relative astrometry can be measured for
the HR 8799 system now spans 16 years.  Here we report on new
epochs of imaging taken between 2009 and 2014, and updated
analysis of these data that has led to improvements in our
astrometric measurements and uncertainties.

\subsection{New Imaging Data}\label{new}

New data was obtained with the Keck II 10 m telescope
with the facility adaptive optics (AO) system \citep{wiz06} and the
near-infrared camera, NIRC2 (PI K. Matthews).  In all observations,
HR 8799A ($V \sim$ 6, \citealt{hog00}) is used as the natural 
guide star for the AO system.  NIRC2 has a plate scale of 9.952 $\pm$
0.002 mas pixel$^{-1}$ and columns that are at a PA of 0.252
$\pm$ 0.009$^o$ relative to North \citep{yelda10}.   Data were taken in both the
$K$-short ($K$s, $\lambda_o$ = 2.146 $\mu$m, $\Delta\lambda$ =
0.311 $\mu$m) and $L$-prime ($L$p, $\lambda_o$ = 3.776 $\mu$m, $\Delta\lambda$ =
0.700 $\mu$m) bands.  NIRC2 is equipped with a wheel
of coronagraphic masks ranging in diameter from 100 to 2000
mas.  When observing at $K$-band we used masks (Table 1) but
none were used for $L$p observations.  

As described in \citet{marois08} and \citet{metchev09}, the data were obtained such
that Angular Differential Imaging (ADI, \citealt{marois06}) processing could be
used during reduction. Observations were therefore conducted in
vertical angle mode, in which the telescope pupil is fixed on
the science camera and the field-of-view (FOV) slowly rotates with
time about the star.  Individual frames of 30 second exposure
time are taken as FOV rotates, ensuring that the PSF of the
planets are not overly ``smeared''.  Because HR 8799 passes
very close to zenith over Maunakea, observations were
generally taken bracketing transit to ensure maximum field
rotation. Sky exposures of the same integration time are taken
separately by nodding several arcseconds away from the star
for the $L$p data.  For $K$s, our reduction method described
in Section \ref{reduc} removes sky background.
Table \ref{tab:log} lists the date of all 
observations, the filter, the size of the coronagraphic
mask, and the total exposure time of all frames.  For
completeness we also list three epochs of previously published observations
on which we performed new analysis for improved astrometry
(see section \ref{reduc}).     

\begin{deluxetable}{lcccc} 
\tabletypesize{\scriptsize} 
\tablewidth{0pt} 
\tablecaption{Log of NIRC2 Observations} 
\tablehead{ 
  \colhead{Date} & \colhead{Filter} & \colhead{Coronagraph} &
  \colhead{Total Int.} & \colhead{Update} \\
  \colhead{(UT)} & \colhead{} & \colhead{Size (mas)} &
  \colhead{Time (s)} & \colhead{or New?} \\
}
\startdata 
2007 Aug 02 & $H$ & 1000 & 3660 & Update\tablenotemark{a} \\
2007 Oct 25 & CH$_{4}$S & none & 2340 & Update \\
2008 Sep 18 & $K$s & 800 & 1160 & Update \\
2009 Jul 30 & $K$p & 600 & 2800 & New \\
2009 Aug 01 & $L$p & none & 3200 & New \\
2009 Nov 01 & $L$p & none & 2250 & New \\
2010 Jul 13 & $K$s & 400 & 1460 & New \\
2010 Oct 30 & $L$p & none & 5900 & New \\
2011 Jul 21 & $K$s & 400 & 3160 & New \\
2012 Jul 22 & $K$s & 400 & 3325 & New \\ 
2012 Oct 26 & $K$s & 400 & 1940 & New \\
2013 Oct 16 & $L$p & none & 1715 & New \\
2014 Jul 17 & $L$p & none & 4900 & New \\
\enddata
\tablenotetext{a}{Originally published in \citet{metchev09}}
\label{tab:log}
\end{deluxetable}

\subsection{Data Reduction and Astrometric Measurements}\label{reduc}

In the years since the initial publication of
\citet{marois08}, a number of systematic biases due to the
method of data collection and 
reduction algorithms being used to derive the astrometry from
Keck have been uncovered and explored.  Specifically, these biases are
introduced by the original implementation of the LOCI
algorithm \citep{lafreniere07} and taking data in ADI mode
\citep{marois06}.  These are summarized in \citet{marois10b}
and include image registration error due to imperfect
knowledge of the star position, PSF elongation due to FOV
rotation, and PSF modification due to self-subtraction.  Newer generation algorithms
that are also based on least-squares but implemented such that
biases are reduced have led to improvements in the
derivation of photometry and astrometry (e.g.,
\citealt{marois10b, soummer12, amara12, meshkat14, fergus14,
  marois14, gg16}).

For this work, we use the Speckle-Optimized Subtraction for
Imaging Exoplanets (SOSIE) algorithm \citep{marois10b}
to both reduce new data obtained between 2010 and 2014, and
re-reduce previously published data from 2007 to
2009\footnote{Keck data were taken by our group in 2004
  but in non-ADI mode.  Because it is non-ADI, only the
  outer two planets are detected, and the new pipeline would
  not improve the astrometry.  Thus, it was not re-reduced
  here, but values from \citet{marois08} are included in the
  subsequent analysis.}.  Briefly, the
algorithm first performs basic reduction of all images (dark
and sky subtraction, flat field reduction, bad pixel removal)
and corrects for NIRC2 distortion using the solution from
\citet{yelda10}.  Next, the images are registered.  For
coronagraphic data, the star is faintly visible through the
occulting spot and can be used for centroiding.  For non-corongraphic data, the core
of the star is saturated, so cross-correlation with a
reference PSF is used.  Then the PSF subtraction is performed
using a least-squares algorithm.  Finally, the images are
rotated such that north is up and the images are combined.

To avoid biasing the astrometry due to the impact of ADI and
LOCI processing on the planet PSF from self-subtraction, a forward modeled PSF template is calculated in all
optimization sections based on the LOCI parameters used.  This PSF is then used to fit for
astrometry and photometry for each planet.  To derive
uncertainties, a model PSF is used to subtract the planet from
the image once the position and flux are known.  The residual
noise in the region surrounding the location of the subtracted
planet is calculated.  We then perform a Monte Carlo
simulation in which we slightly vary the position of the
planet center from the derived location, subtract the PSF, and
again compute the noise.  This allows us to generate a ``noise
curve'' as a function of planet position.  In order to
calculate the one-sigma positional uncertainties, we determine the
offset that yields an increase in the local noise
by a factor of a square root of 2.  This is done separately
in the X and Y direction, and the uncertainties in each
direction are averaged to derive a final positional uncertainty.  A similar process is performed for
photometry.  Further details on our method of derivation of
uncertainties is described in \citet{marois10b} and \citet{galicher11}.

In the course of obtaining our observations, we also noted that an
uncertainty was also introduced by the NIRC2 coronagraphic masks.
Specifically, while the semi-transparent nature of these masks
is useful for obtaining unsaturated centroids for the star, we
found a systematic shift in source
positions when the focal plane mask is in place versus when it
is not in place.  Through a series of tests with the Keck AO
internal fiber-fed point source, we determined the extent of this
offset by marching the source across the NIRC2 FOV.  We found
that for the most part, this is a uniform shift
across the FOV - this was by design, as the focal plane mask
was given a slight tilt of 2.603 $\pm$ 0.003$^{\circ}$ (K. Matthews, private
communication).  Thus, for relative astrometry, there should
be no impact from having the focal plane mask in the optical path.  However, for point sources under
the coronagraphic spot, the shift was found to be slightly
less than outside of the spot (on the order of $\sim$0.1 pixels), therefore impacting our
relative astrometry.  The cause for this difference is
unknown.  To measure this offset on each data set, an
iterative ``rotation axis'' technique was developed to search
for the optimal rotation axis that maximizes the SNR of the
planets.  Using this, we find that the uncertainty on the star
position is now typical on the order of $\sim$2 mas.

The resulting X and Y positions and uncertainties were
converted into arcseconds using the NIRC2 plate scale from
\citet{yelda10}, and the uncertainties in the plate scale and
north angle offset are added in quadrature with the positional uncertainties.  All values for all four planets are given in
Table \ref{tab:astr}.  The final astrometric uncertainties range from 3 to 22 milliarcseconds.

\section{Orbit Fitting}\label{orbits}

\subsection{The Construction of a Self-Consistent Data Set}\label{construction}

Fitting relative astrometric orbits to data on long-period objects, while
technically straightforward, can lead to biased
results.  Fitting
relative orbits to a small percentage of a $>$50 year orbit tends
to yield a preference for periastron passage close to the
epoch during which the data was taken.  As a result,
artifically high eccentricities tend to also be preferred.
Furthermore, systematics between astrometric data taken from 
different cameras are noticeable and impactful when
full phase coverage is obtained - the situation is worse when
only a tiny fraction of the orbit has been measured.  In the
case of directly imaged planets, a further complication is
introduced by the choice of algorithm used to enhance the
contrast and yield a robust detection of the planet. Each
algorithm uses different methods of deriving astrometry, which can
further skew resulting orbital parameters.  Assessment of the biases introduced by
each pipeline, such as those discussed in relation to SOSIE in
\citet{marois10b}, are ongoing (e.g., \citealt{amara12,
  pueyo15}).  Even when all possible biases are
accounted for and uncertainties enlarged to attempt to
encompass these errors, systematics across multiple data sets
remain.  As an example, it has been noted by multiple authors
that the previously published Keck astrometry on HR 8799
\citep{marois08,marois10a}, when combined with other data,
yields poor 
$\chi^2$.  For instance, fitting orbits to our previous astrometry for HR
8799b yields a best-fit reduced $\chi^2$ of 2.5.  The probability of obtaining this value for
$\chi^2$ under the assumption of Gaussian uncertainties is
0.002\%, highlighting the remaining systematics in our data
set. Our improvements in reduction and derivation of astrometry
improve the overall $\chi^2$ fits by factors of $\sim$4-8,
with associated probabilities of Gaussian uncertainties $>$90\%.

Still, even with our improved reduction, systematics between our 
dataset and data presented in the literature (from multiple
cameras and reduction pipelines) remain.  The Keck dataset
is by far the most extensive, spanning $\thicksim$10 years for HR 8799b
and c, $\thicksim$7 for HR 8799d, and $\thicksim$5 for HR 8799e.  Thus, we elect
here to estimate the orbital parameters for the system using
\textit{only} this data set.  This circumvents most of the
remaining systematics, though it does lower the overall time
baseline by not including the HST data points from \citet{lafreniere09} or \citet{soummer11}.
However, we conducted several sample orbit fits for HR
8799b, c, and d to determine the
impact of the HST points on solutions.  Due to the large
uncertainties on the HST astrometry, all
orbital parameters obtained with and without the HST data were
consistent to within their uncertainties.  The Keck data set,
with among the smallest uncertainties obtained, remains the
biggest weight in all fits.  We also only utilize data reduced
through our reduction pipeline, electing not
to use the results from \citet{currie12} or
\citet{currie14}.    

\subsection{Acceleration Detections}\label{accel}

\begin{figure*}
\epsscale{0.8}
\plotone{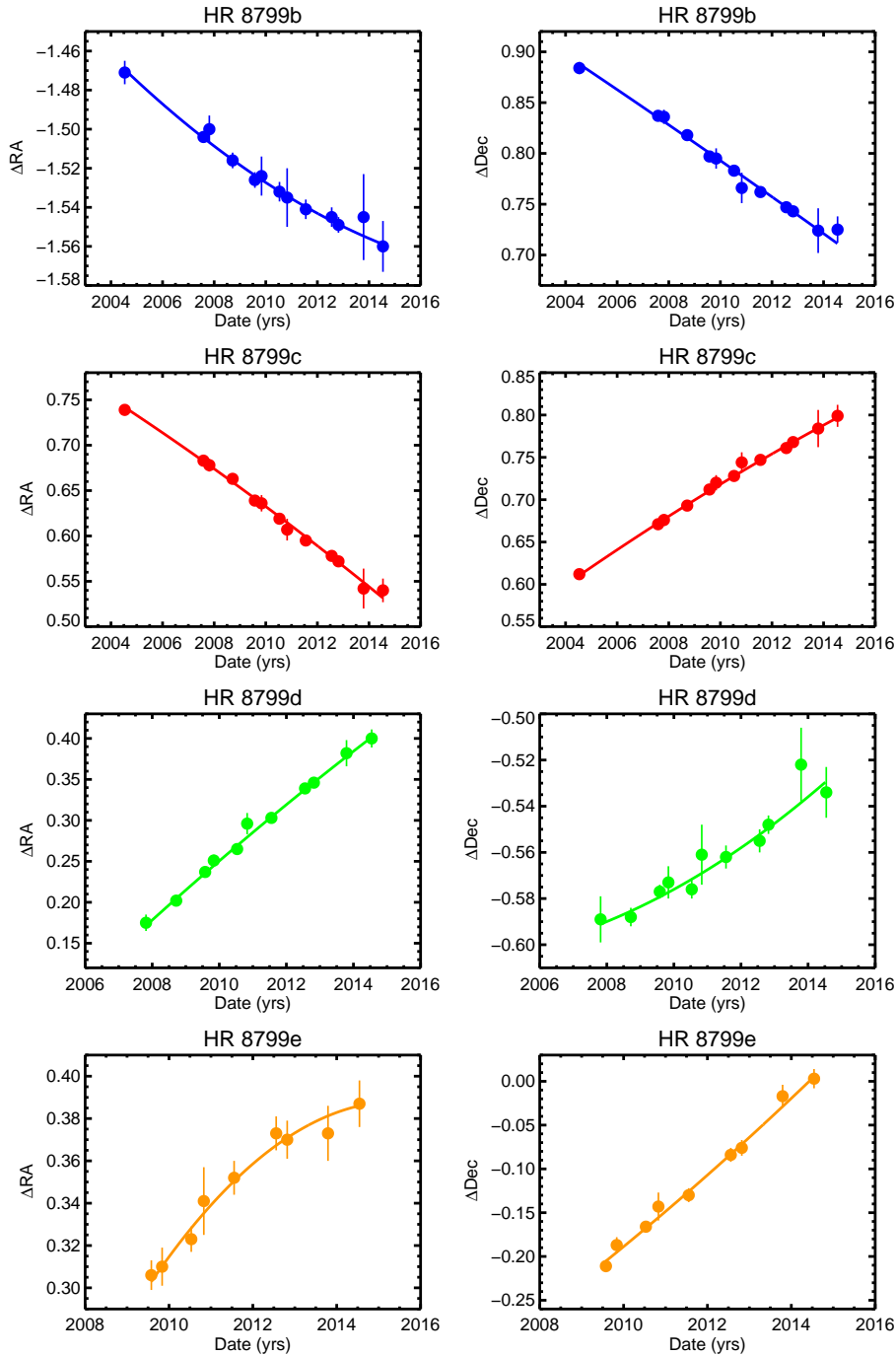}
\caption{Keck astrometric measurements as a function of time
  for all HR 8799 planets.  We use the same color scheme
  throughout to represent each planet: HR 8799b = blue, HR
  8799c = red, HR 8799d = green, and HR 8799e = orange.  Overplotted are polynomial fits to
the data used to determine whether acceleration was detected
\citep{jlu09}.  Statistically significant curvature is
detected with Keck data only for HR 8799b and e, but has not
yet been detected for HR 8799c and d.  We anticipate that with
continued monitoring and similar error bars, we will detect acceleration using Keck
data in 2016 for HR 8799c and d.}
\label{fig:astr_poly}
\end{figure*}

With our improved astrometric data, we can first
determine whether any orbital curvature (or acceleration) has
been detected.  In Figure \ref{fig:astr_poly}, we plot the x (RA) and y (Dec)
astrometry from Keck used in our 
analysis as a function of time for all four planets.  In order
to determine whether 
acceleration was detected for any of the planets, we
fit second-order polynomials of the form given in Equations 3
and 4 of \citet{jlu09} to the data shown in Figure
\ref{fig:astr_poly}.  These polynomials are shown as lines overplotted on the
data in Figure \ref{fig:astr_poly}.  This gives us an estimate of the velocity
and acceleration in x and y and an uncertainty.  We then
converted these into radial and tangential components of
velocity and acceleration by a transformation from cartesian
to spherical coordinates.  Since true orbital acceleration
will only have a negative radial component, we consider
orbital acceleration ``detected'' if the 
measured radial acceleration plus three times its uncertainty
is less than zero (or effectively that it is detected to
3$\sigma$).  This is analogous to the procedure for
determining the significance of an acceleration detection
described in \citet{jlu09}   

Using this method, we find that we have detected acceleration
to 3.4$\sigma$ for HR 8799b (-0.35 $\pm$ 0.10 mas/yr$^{2}$)
and to 5.4$\sigma$ for HR 8799e (-2.55 $\pm$ 0.48 mas/yr$^{2}$).
Acceleration is not detected significantly for HR 8799c
(-0.32 $\pm$ 0.21 mas/yr$^{2}$, 2.1$\sigma$) or HR
8799d (-0.65 $\pm$ 0.56 mas/yr$^{2}$, 1.2$\sigma$).  For all four planets, the (non-physical) tangential acceleration
derived from the fits is insignificant ($<$0.5$\sigma$).  We anticipate that
with continued measurements at a similar cadence to our
current observations (one or two times per year), orbital acceleration will be detectable for in
2016 for both HR 8799c and HR 8799d using Keck data alone
(note that including the HST datapoint from 1998 would yield
an acceleration detection for these planets). 

\subsection{Orbital Fitting}\label{orbfit}

\begin{figure*}
\epsscale{1.0}
\plotone{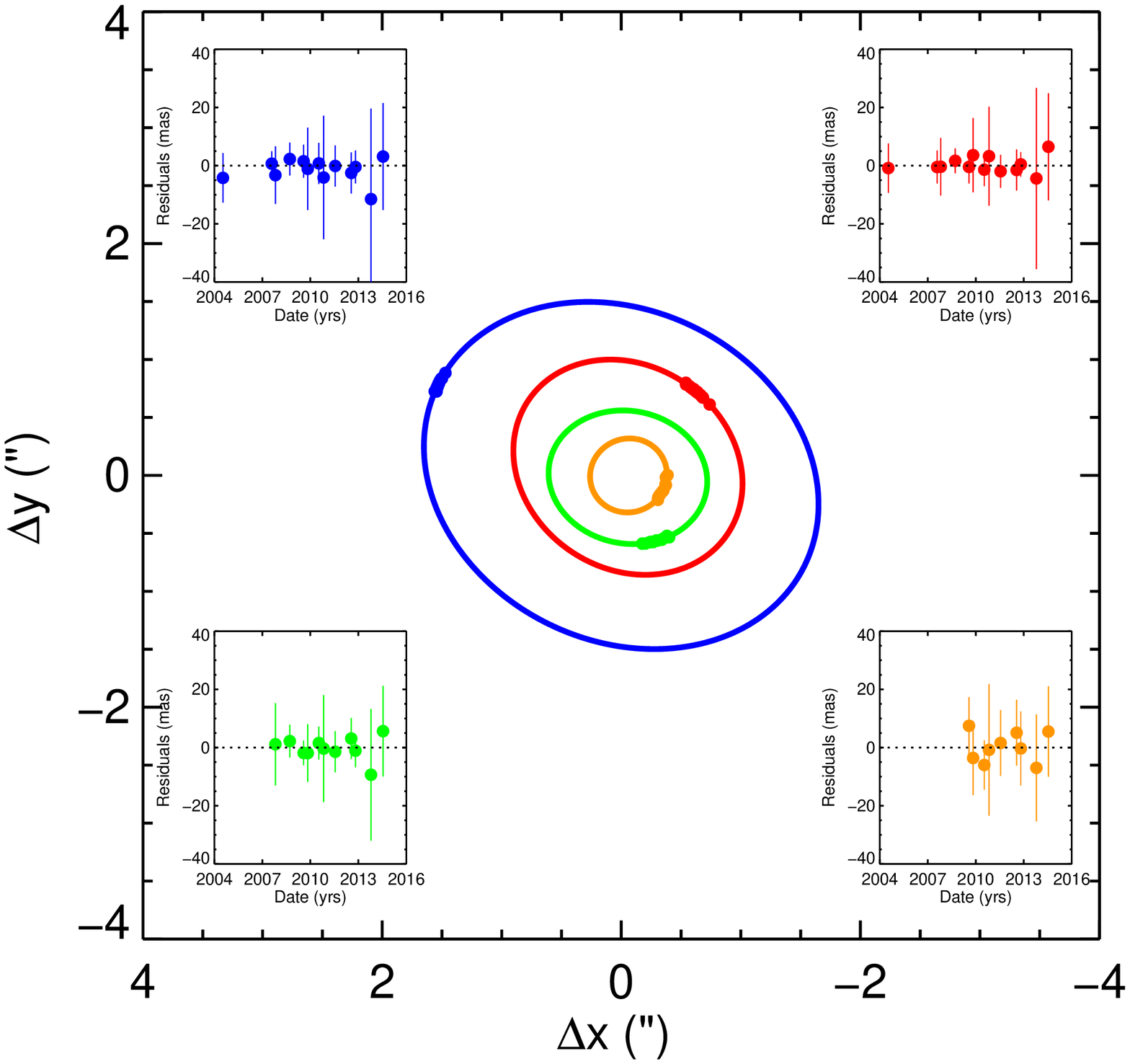}
\caption{Sample low eccentricity orbits that are consistent
  with our astrometry.  Each orbit shown is within 1$\sigma$
  of the best fit solution.  The side panels shown the size of
the residuals to each of these fits.  All residuals are
consistent with zero to within our uncertainties.  For all
four planets, our best fit reduced $\chi^{2}$ is less than 1,
suggesting that our uncertainties are slightly overestimated.
This is likely due to our conservative approach to uncertainty
assignment for both the positions of the planets and the
positions of the central star.}
\label{fig:orbs_inset}
\end{figure*}

\begin{figure*}
\epsscale{1.0}
\plottwo{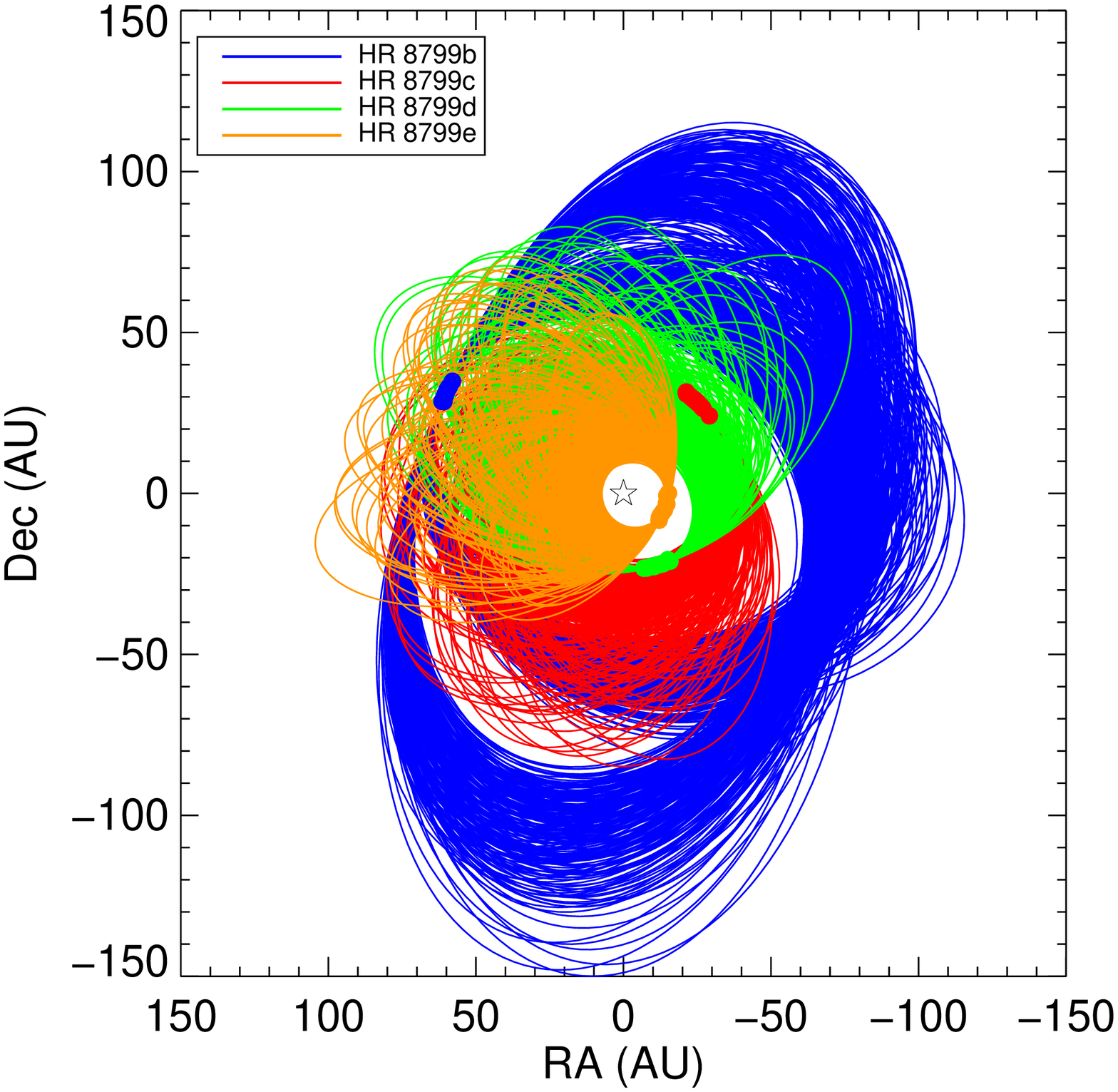}{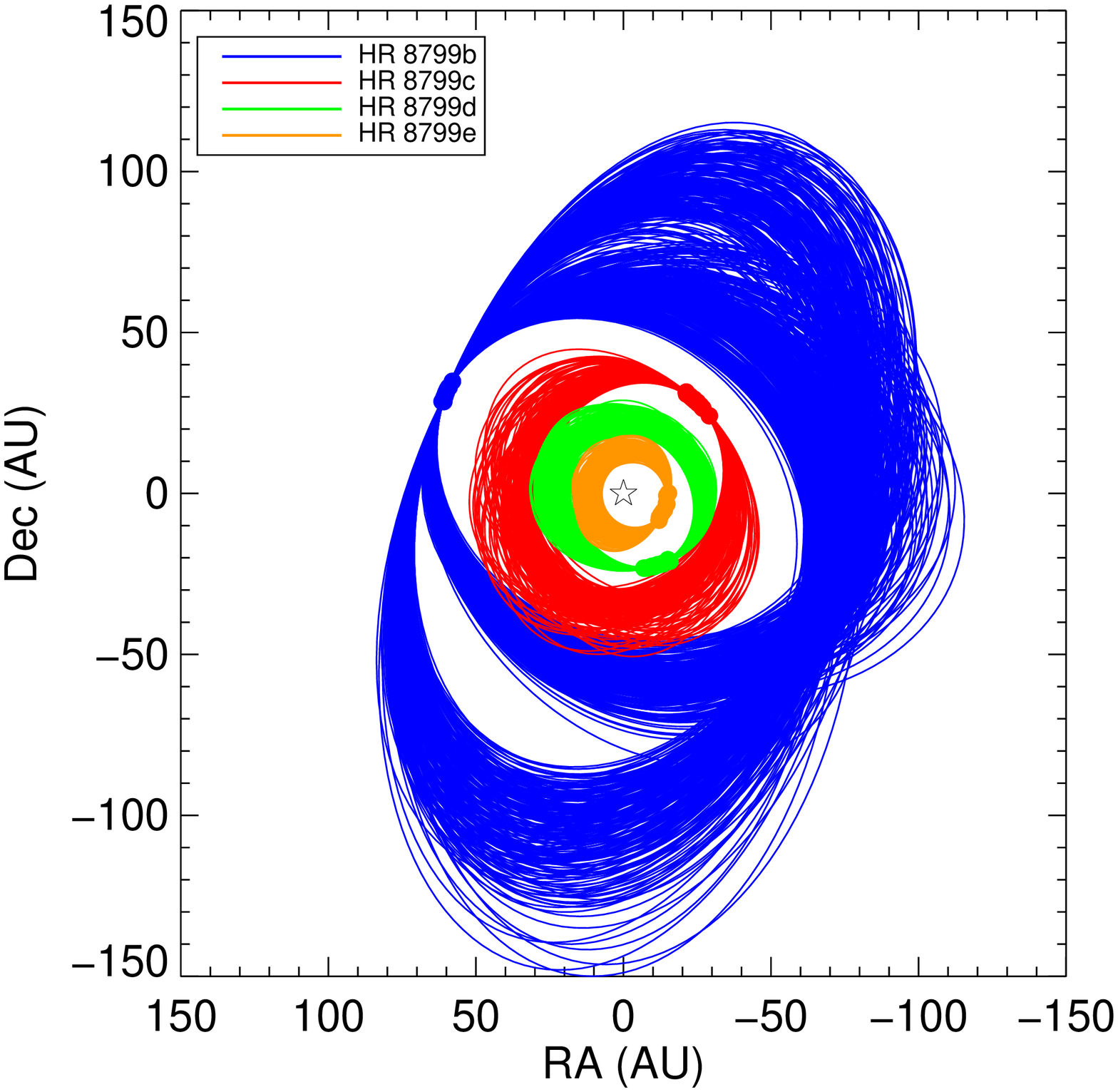}
\caption{\textbf{Left}: A subset of randomly selected orbits
  for each planet that are within 3$\sigma$ of the best fit
  solution.  Crossing orbits are still allowed by the data,
  while stability requirements would suggest that such orbits
  are unlikely to be the true orbits of the planets.
  \textbf{Right}: A subset of randomly selected orbits where
  the Hill radius of the planet does not come within the Hill radius of any other planets.  The
distribution of orbital parameters for these ``non-crossing''
orbits are shown in Figure \ref{fig:nocross_1d}.} 
\label{fig:orb_fam}
\end{figure*}

Using the astrometric measurements described above, we
now seek to determine the currently allowed astrometric orbital parameters for the
HR 8799 planets.  Our model
for the relative orbit always contains six free parameters:
period (P) (related to the semi-major axis (a) through the
total system mass),
eccentricity (e), time of periapse passage (T$_o$), inclination (i), position
angle of the ascending node ($\Omega$), and longitude of periapse
passage ($\omega$).  We fix the mass of the star to that
measured in \citet{baines12}, 1.516  M$_{\odot}$ for an
age of $\thicksim$30 Myr, and increase the uncertainties on
this value from the published numbers to $\pm$0.15 M$_{\odot}$
 (or 10$\%$ of the mass) to account
for any possible additional uncertainties in the evolutionary
models used to derive the mass.  In order to convert from on-sky to
physical units, we use the distance to the star measured by \citet{van07}
of 39.4 $\pm$ 1.1 pc.

Initial orbital parameters ranges are first estimated by
mapping $\chi^2$ through a Monte Carlo minimization routine
described in \citet{ghez08}.  All orbital parameters are
sampled from uniform distributions except for inclination,
which is sampled from a distribution uniform in cos(i).  Note
that because such a small 
fraction of the orbit has been mapped for these planets
($\thicksim$3$\%$ for HR 8799b, $\thicksim$6$\%$ for HR 8799c, $\thicksim$12$\%$ for HR
8799d, and $\thicksim$6$\%$ for HR 8799e), the ``best fitting'' orbit is
not strictly meaningful in its own right.  However, the
value of reduced $\chi^2$ we obtain for the best fits is a
gauge of the fidelity of our astrometry and error bars, and a
check for systematics remaining in our Keck-only dataset.  For
our choice of data, the best fit reduced $\chi^2$ values are 0.36, 0.29, 0.52,
and 0.44 for HR 8799b, c, d, and e, respectively, implying that
our uncertainties are in fact slightly overestimated.  This is
likely due to our conservative approach in assigning an
uncertainties, including assumptions about the star centroid
and using our Monte Carlo simulation to derive the uncertainties
in the planet position.  However, it also shows that our
method is likely incorporating any remaining systematic
uncertainties and validates the choice of using a single
consistently reduced data set.  As an
example of the quality of the fits, we plot low-eccentricity solutions that are
consistent with the astrometry in Figure \ref{fig:orbs_inset}.
Positional residuals between these fits and the astrometry are
shown in the insets for each planet.  All data points are
consistent with these fits to within their 1$\sigma$ uncertainties.

Once we broadly determined the range of allowed parameters, we
mapped their probability distributions using a Monte Carlo
simulation.  First, 100,000 artificial data sets are
generated to match the observed data set in number of points,
where the value of each point (including the distance and mass
of the star) is assigned by randomly drawing
from a Gaussian distribution centered on the best-fit value with a
width corresponding to the uncertainty on that value.  Each of
these artificial data sets is then fit with an orbit model as
described above.   From these trials, we saved the best fit
model.  Our orbital analysis approach is somewhat different from
other works, relying on a more standard Monte Carlo
implementation than on methods like MCMC.  The analysis
performed here is computationally intensive, taking anywhere 
from a factor of 1.5-2.5 times longer to run on an equivalent
data set that converges, but also avoids potential issues such as getting stuck in
local minima, forcibly exploring parameter space.  Methods
that are less likely to have issues with complex distributions
are preferable in situations such as ours where this could be concern, in spite of the
computational expense.  Given proper implementation and
exploration of parameter space, the  results from the two
algorithmic implementations  should return 
equivalent results and uncertainties 
(e.g., \citealt{derosa15}).

For HR 8799e, we also computed the best fit obtained
after an additional weighting for the ``likelihood'' of a given orbit.
As discussed in Section \ref{construction}, highly undersampled orbits are often ``best-fit'' by solutions
that are biased toward high eccentricities and T$_{o}$ near
the time the data is taken, even when astrometric systematics
are minimized.  This bias impacts the HR 8799e simulations,
where the phase coverage is the 
smallest and the astrometric uncertainties the greatest.  The
difference between the distributions for the other planets
whether weighted or not is small.  We therefore applied a correction
factor to the HR 8799e fits, which we calculated by dividing
the time baseline of our 
observations by the average time the planet would spend in
another portion of its orbit with comparable arclength, or
distance traveled.  We multiplied each $\chi^2$ by this
correction factor, and then saved the best fit after applying
this correction.  In order to verify that this correction
factor was appropriate, we tested this process on simulated
astrometry sampled from a notional orbit for HR 8799e that had
an eccentricity of $\sim$0.05 and went through periastron
passage in 1985.  We assigned this astrometry equivalent error
bars to our actual data.  In this test, we found that when we
did not weight our $\chi^2$ values,  25\% of the solutions had T$_{o}$ coincident with the time the
data was taken, with 75\% within $\pm$5
years of the time the data was taken.  With weighting, this
fell to 0.3\% of solutions with T$_{o}$ during the time the data was taken and 1\%
within 5 years of the data.  In addition, without weighting,
64\% of solutions yielded eccentricity $>$0.5, whereas 0.2\%
of weighted solutions had eccentricities this high.  Thus we
believe that in the case of HR 8799e, the distribution of
parameters for weighted orbits has a much greater likelihood
of encompassing the true orbital parameters.  Note that the
impact of  weighting on solutions where HR 8799e is
currently at apoastron with a high eccentricity and thus
short orbital period are minimally impacted by this weighting,
as the probability of observing an eccentric planet at
apoastron is high (though these solutions are also unlikely,
see Section \ref{orbparm}).

In order to verify that no data points in our Keck
data set given in Table \ref{tab:astr} were significantly
impacting our resulting orbital parameters,
we performed the same fitting routine with each data point
individually removed.  For HR 8799c and e, the exclusion of
any datapoint has no impact on the parameter ranges.  For HR
8799b and d, the removal of the 2008 datapoint has a modest
impact on the range of periods (and therefore
SMA), increasing the upper limit on periods by about 10$\%$ in
both case.  The same is true for HR 8799b in the case of the
2004 data point.  In the case of the 2008 data, the points are
the only ones for that year and have small error bars.  For
2004, the error bars are larger but the datapoint is unique in
time sampling.  We also see with the removal of the 2004 data
point that the increase in allowed periods corresponds to a
few degree increase in the range of inclinations for
HR 8799b.  Still, given that the constraints on all parameters
are quite broad using the full dataset, we assert that the
impact of any one data point on our fits is very modest.

We note that our goal is to obtain all orbital
parameters currently allowed by our astrometry.  While other
authors have focused on solutions based on stability
criterion (e.g., \citealt{soummer11, maire15, zurlo15}), we were interested in what regions of
parameter space could be ruled out by the self-consistent data
set alone assuming Keplerian orbits.  Future work will include assessing the stability of orbits allowed
from the Keck data.

\subsection{Allowed Orbital Parameters}\label{orbparm}

To demonstrate the allowed X-Y phase space of orbits by the
astrometric data, we plot in Figure \ref{fig:orb_fam} the
allowed orbits for each planet (converted to AU using a
distance of 39.4 pc).  The left hand panel
demonstrates all phase space, while the right hand
panel shows only the two-dimensional projection of orbits that
do not cross the Hill radius of another planet. 

\begin{figure*}
\epsscale{0.9}
\plotone{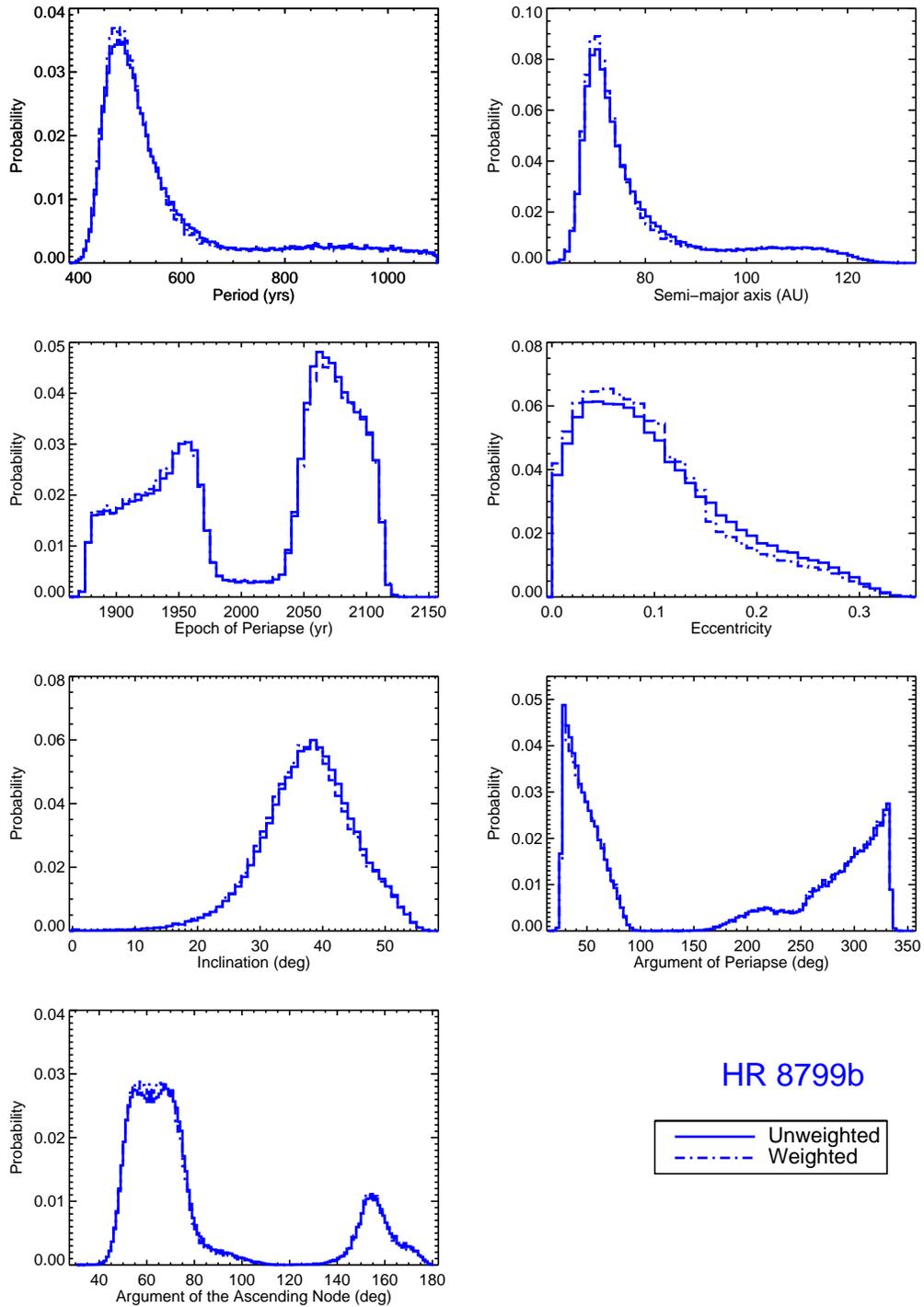}
\caption{Results from our Monte Carlo simulation for HR
  8799b.  One-dimensional PDFs from all possible solutions are shown for each
  of the seven free parameters. The solid lines represent the
  solutions that are unweighted by the likelihood of catching the
  planet in its current phase, while the dashed line shows
  solutions weighted by this likelihood.  Low eccentricity solutions
  are favored, as is an inclination of 38 $\pm$ 7$^{\circ}$.}
\label{fig:1d_b}
\end{figure*}

\begin{figure*}
\epsscale{0.9}
\plotone{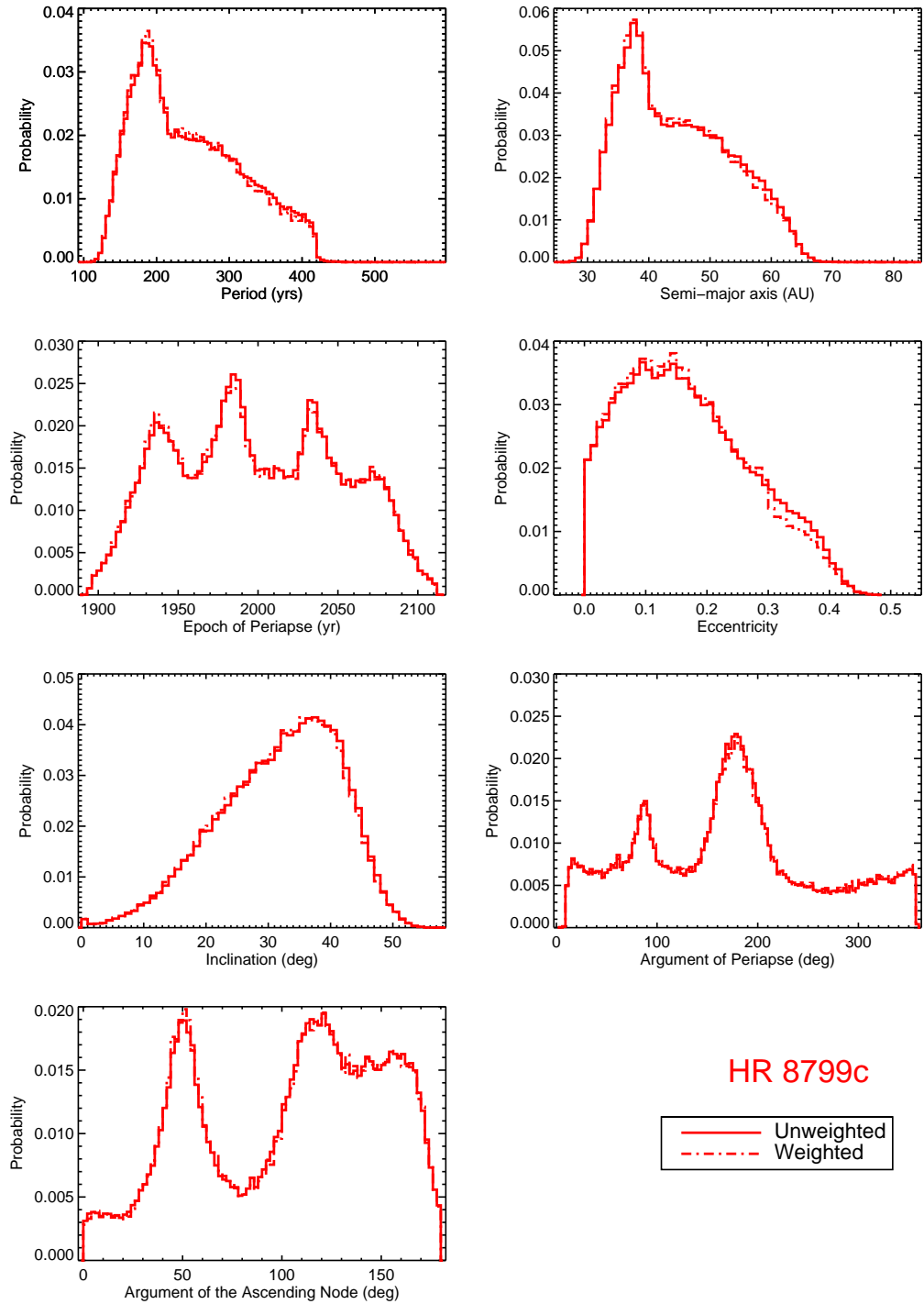}
\caption{Results from our Monte Carlo simulations for HR
  8799c.  One-dimensional PDFs from all possible solutions are shown for each
  of the seven free parameters.  Solid lines represent
  unweighted solutions while dashed lines show weighted solutions.  All allowed eccentricities
  have a value of $<$0.5, and the preferred inclination is 37
  $\pm$ 12$^{\circ}$.}
\label{fig:1d_c}
\end{figure*}

\begin{figure*}
\epsscale{0.9}
\plotone{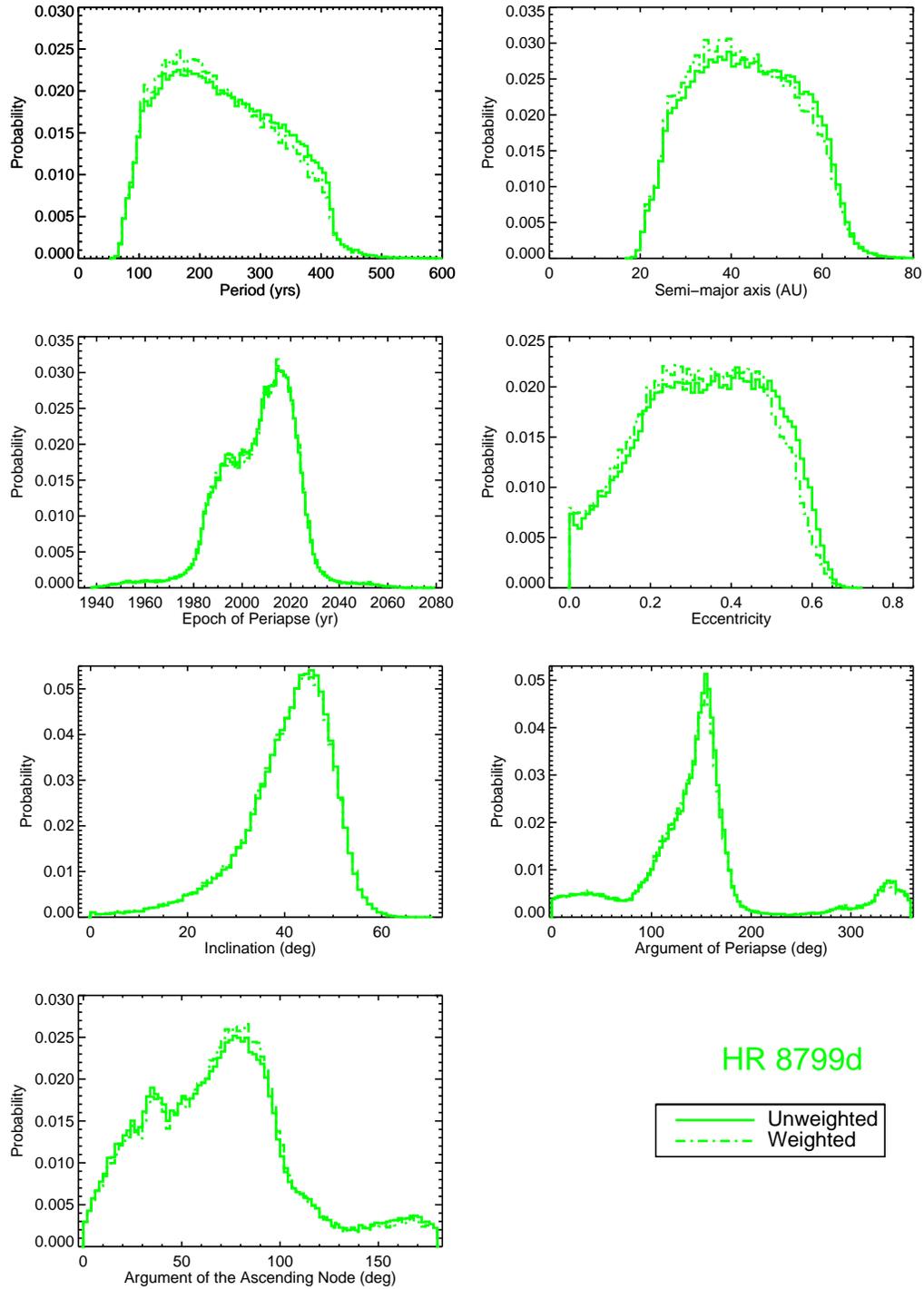}
\caption{Results from our Monte Carlo simulations for HR
  8799d.  One-dimensional PDFs from all possible solutions are shown for each
  of the seven free parameters.  Solid lines represent
  unweighted solutions while dashed lines show weighted solutions.  An eccentricity upper limit
  of $<$0.7 is seen, with an inclination of 45 $\pm$
  8$^{\circ}$.  While this inclination is slightly offset from HR 8799b
  and c, it is consistent to within the uncertainties.}
\label{fig:1d_d}
\end{figure*}

\begin{figure*}
\epsscale{0.8}
\plotone{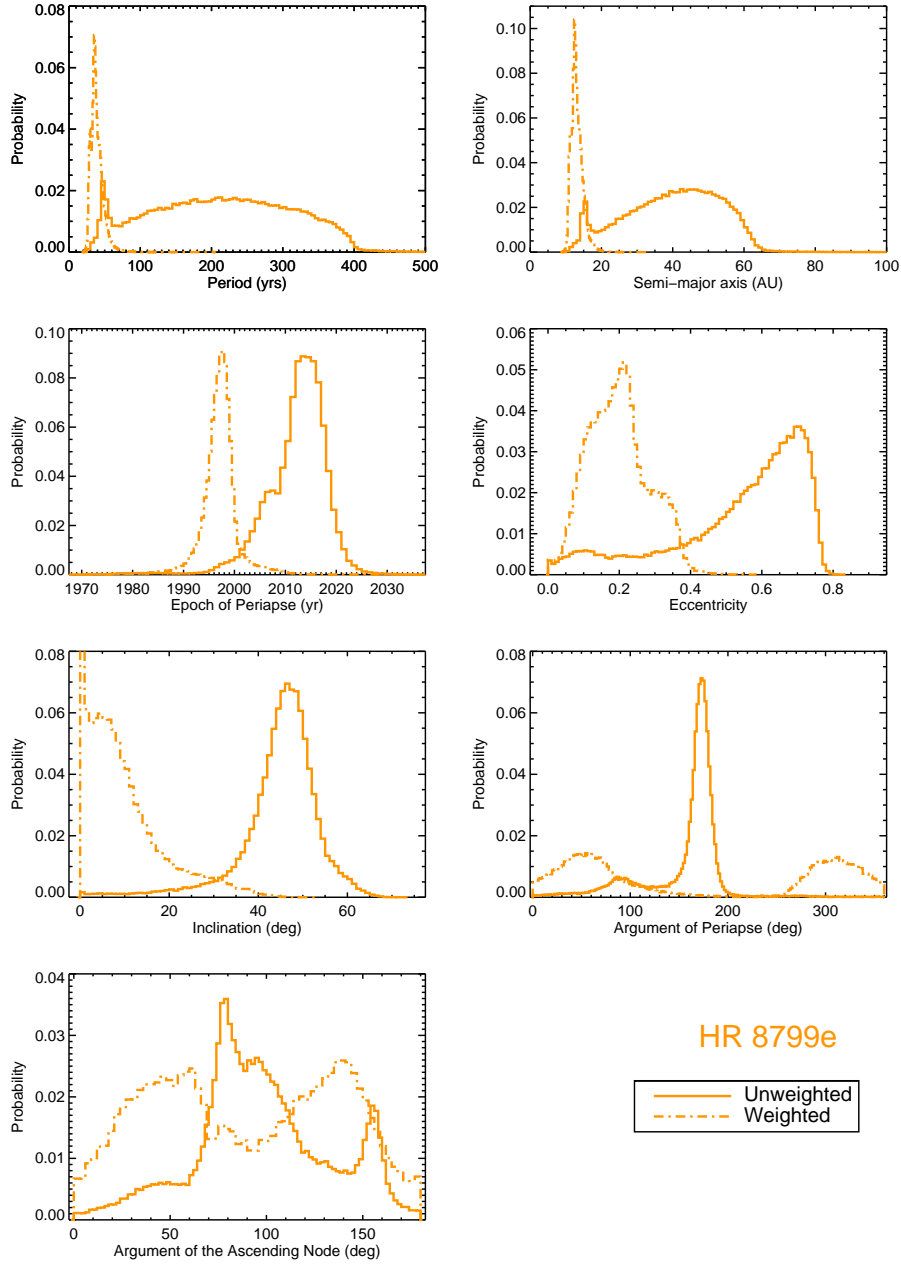}
\caption{Results from our Monte Carlo simulations for HR
  8799e.  One-dimensional PDFs from all possible solutions are shown for each
  of the seven free parameters. The solid lines represent the
  solutions unweighted by the likelihood of catching the
  planet in its current phase, while the dashed line shows
  solutions weighted by this likelihood.  High eccentricities
  are favored in the unweighted case, along with periastron
  passage close to the current epoch.  The inclination also
  shifts from about 50$^{\circ}$ in the unweighted case to
  preferring face on in the weighted case.  Because of the
  higher astrometric uncertainties, our constraints on the
  orbital parameters of this planets are still quite limited.}
\label{fig:1d_e}
\end{figure*}

The full distributions of orbital parameters are shown in
Figures \ref{fig:1d_b} through \ref{fig:1d_e}.  For all planets, we present both
``weighted'' and ``unweighted'' solutions, though they only
differ significantly for HR 8799e.    

\begin{figure*}
\epsscale{1.0}
\plottwo{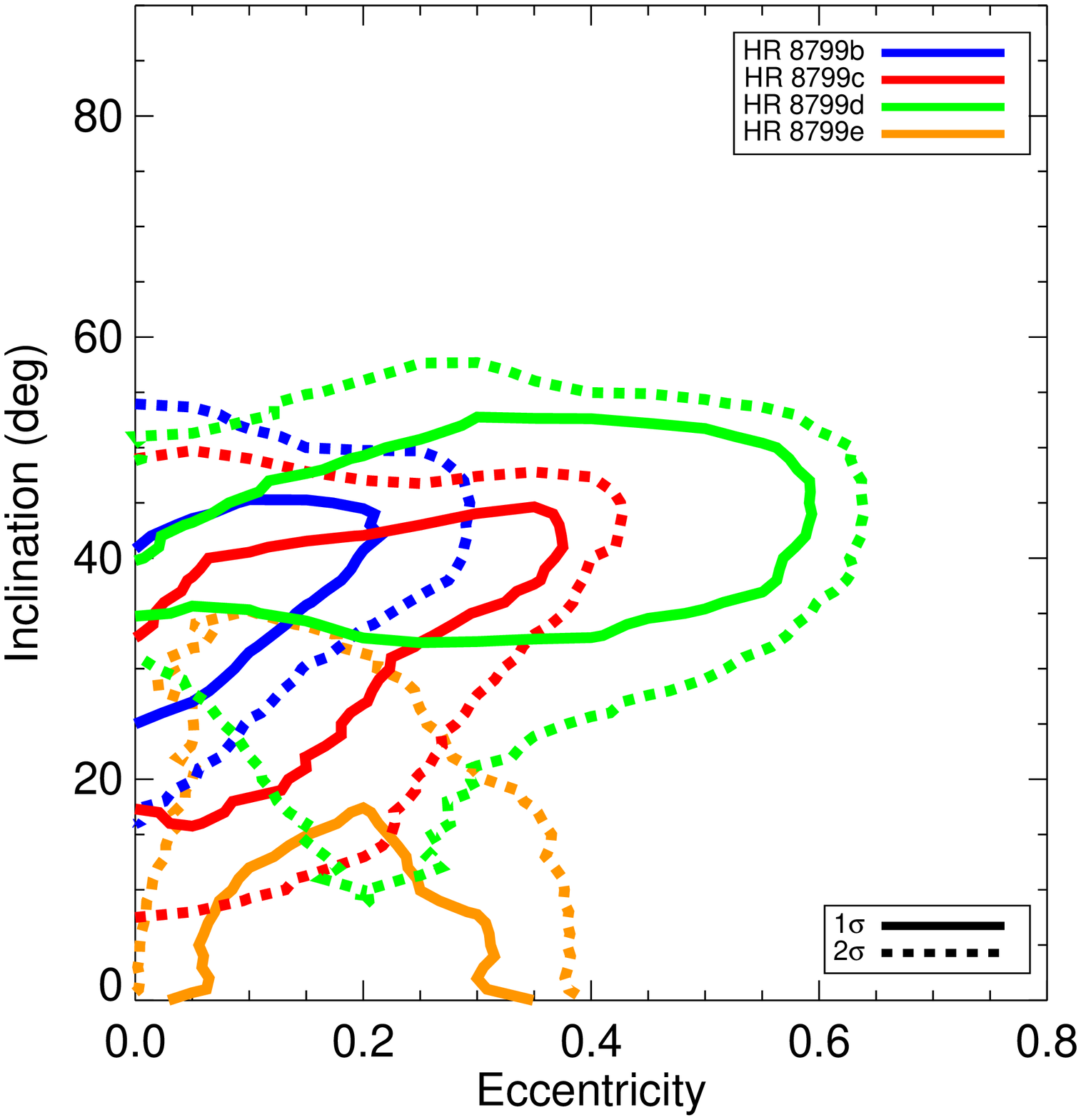}{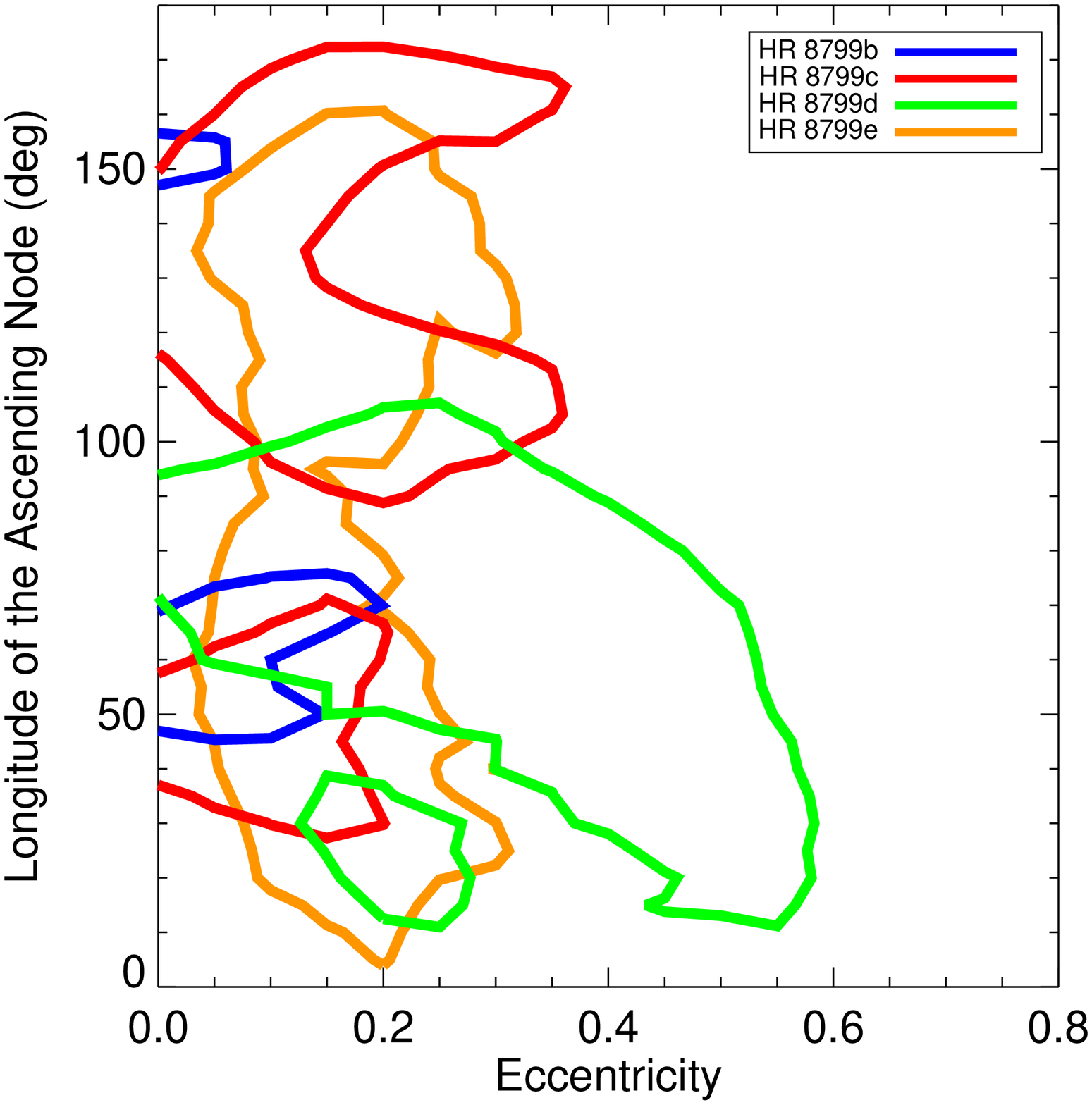}
\caption{\textbf{Left}: The joint-PDF between eccentricity and
  inclination, shown as 1- and 2-$\sigma$ contours for each
  planet (using the weighted fits for HR 8799e).  As stability requirements likely necessitate low
  eccentricity orbits, we note that and inclination of
  $\sim$30$^{\circ}$ is allowed for HR 8799b, c, and d at low
  eccentricity at 1-$\sigma$ (solid line), and for HR 8799 e
  at 2-$\sigma$ (dashed line).  
  \textbf{Right}: The joint-PDF between eccentricity and
  longitude of the ascending node ($\Omega$), shown as
  1-$\sigma$ contours (2-$\sigma$ is not shown to enhance
  clarity).  At low eccentricity, all planets are consistent
  between 50-70$^{\circ}$.} 
\label{fig:2d_contour}
\end{figure*}

\begin{figure*}
\epsscale{0.8}
\plotone{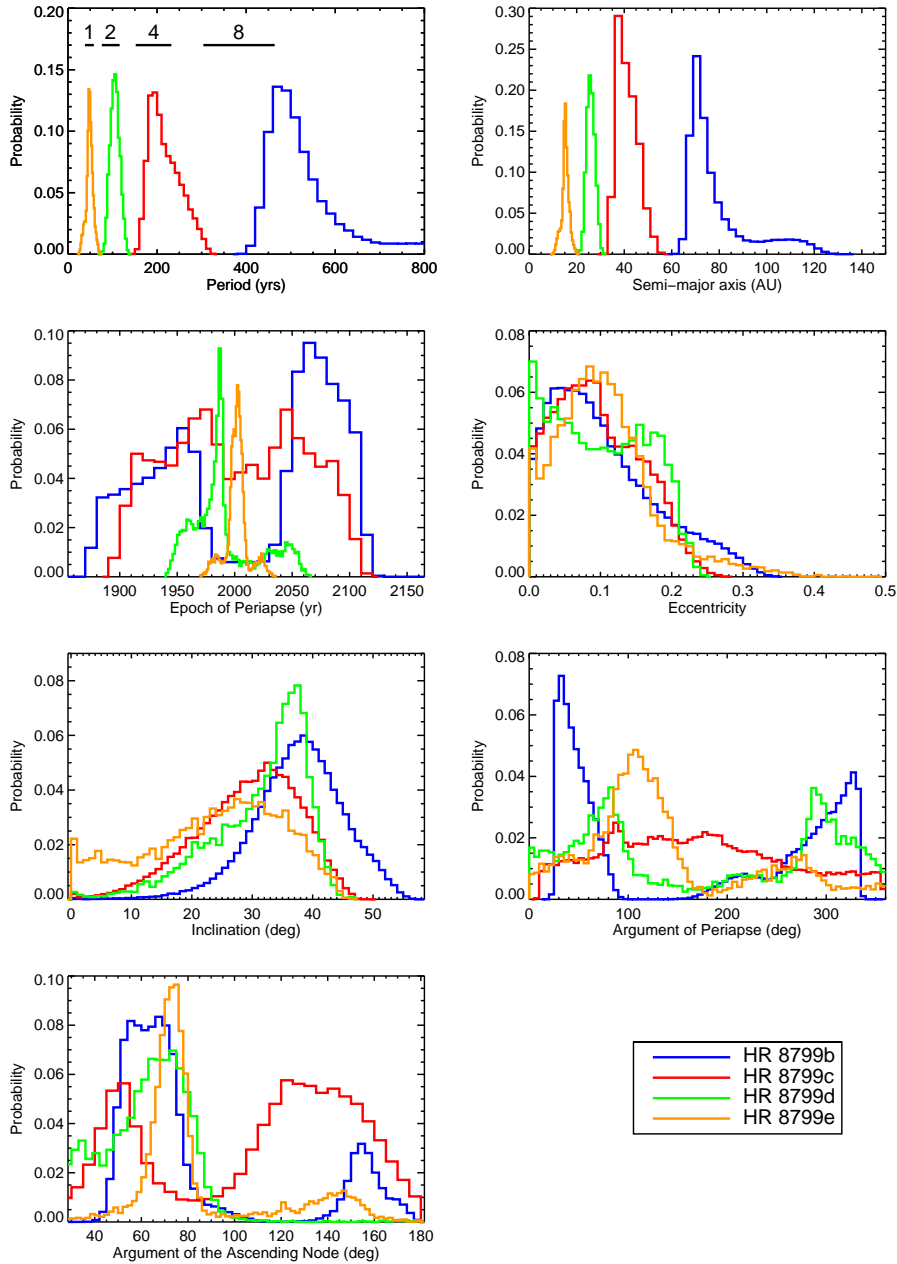}
\caption{One-dimesonsional PDFs for all four planets for only
  those orbits that do not come within the Hill radius of each
  other.  The solutions for HR 8799e are drawn from weighted simulations.  Eccentricity upper limits are 
  $<$0.3 for all planets, and orbital plane
  distributions remain fully consistent, with strong overlap
  in both inclination and $\Omega$.  Also noted is the
  locations of 1:2:4:8 period ratios assuming the peak
 and 1$\sigma$ values of the distribution for HR 8799e is the start of the chain.  In
  this case, all four planets are consistent with these ratios
to within 1$\sigma$.}
\label{fig:nocross_1d}
\end{figure*}

\begin{figure*}
\epsscale{0.8}
\plotone{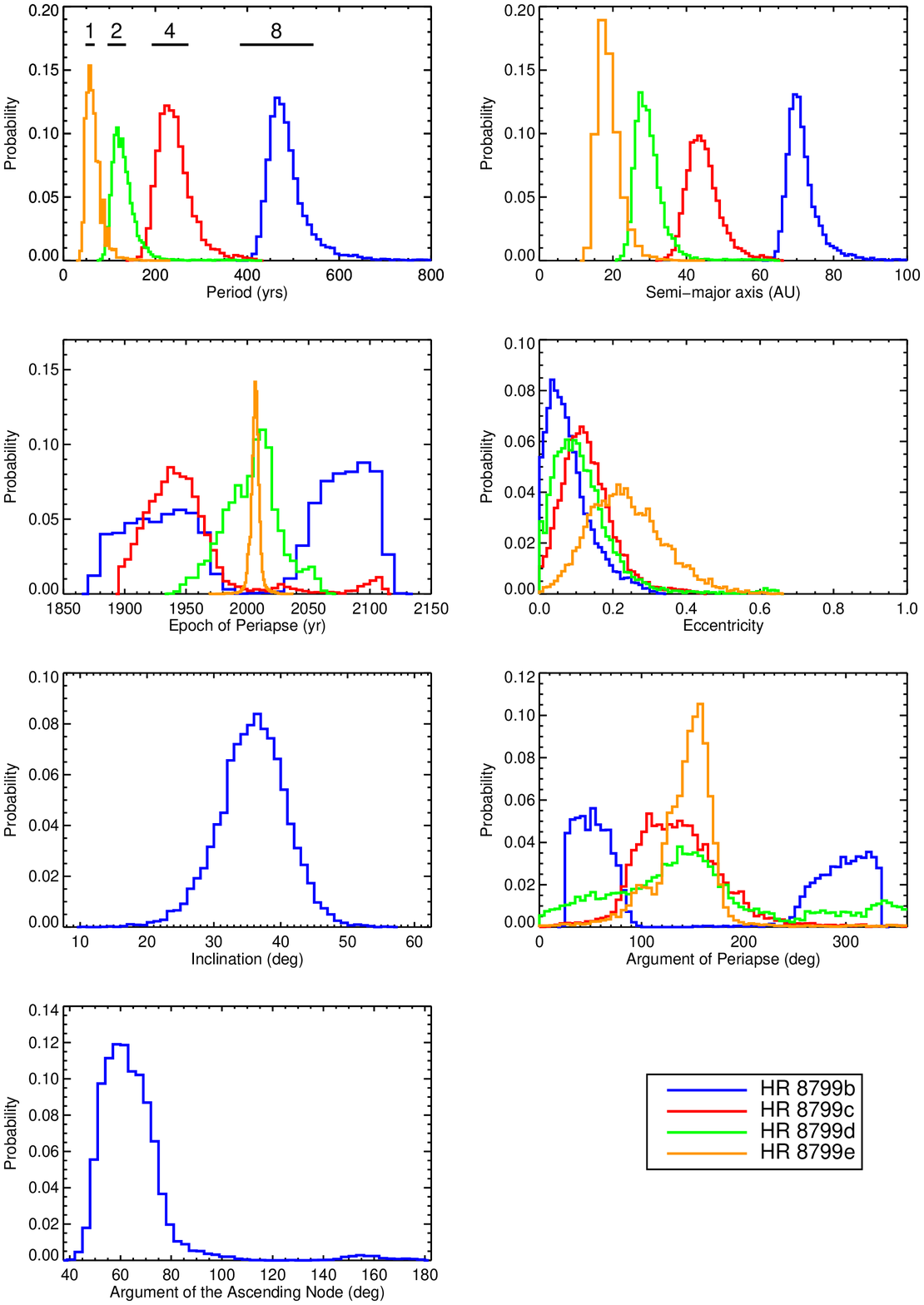}
\caption{One-dimesonsional PDFs for all four planets under the
assumption that they are coplanar.  Inclination and $\Omega$
are sampled from 10000 randomly selected solutions for HR
8799b in our larger Monte Carlo simulation.  Assuming
coplanarity with HR 8799b leads to eccentricity upper limits
of 0.3 for HR 8799c and d, and 0.5 for HR 8799e.  The reduced
$\chi^2$ for these fits for HR 8799c, d, and e are small,
showing that coplanarity is completely consistent
with our data.  Also, as in Figure \ref{fig:nocross_1d}, we note the
  locations of 1:2:4:8 period ratios assuming the
  distribution for HR 8799e is the start of the chain.  All four planets align very well with these
  period ratios.}
\label{fig:mutinc_1d}
\end{figure*}

The strongest orbital constraints are found for HR 8799b
(Figure \ref{fig:1d_b}).  Strong eccentricity
upper limits can be placed at $<$0.3.  Generally
face-on orbits are not allowed, and we find a
preferred inclination of 38 $\pm$ 7$^{\circ}$.  For the other
orbital plane parameter, argument of the ascending node ($\Omega$), we find a peak near
$\sim$65$^{\circ}$, with a second set of solutions near
$\sim$150$^{\circ}$.  The latter appear to be slightly more
correlated with higher eccentricity solutions.  The peak
of the period distribution is at $\sim$470 years, which implies a peak SMA of $\sim$70 AU.
The other two parameters are largely unconstrained.

For HR 8799c (Figure \ref{fig:1d_c}), we
find a peak at fairly low periods and SMAs, with a broad tail
to higher values.  This corresponds to the tail of
eccentricities that are still allowed, with an upper limit of $<$0.5.
The peak value for period is $\sim$185 years, corresponding to
an SMA of $\sim$38 AU.  The preferred inclination is 37 $\pm$
12$^{\circ}$, consistent with the inclination found for HR 8799b.
Few constraints are found for $\omega$, $\Omega$, or T$_o$.  A
higher number of low eccentricity solutions do favor an
$\Omega$ near $\sim$50$^{\circ}$.

For HR 8799d (Figure \ref{fig:1d_d}), the
lower astrometric precision leads to an
essentially flat eccentricity distribution out to 0.6.
Like HR 8799c, the period distribution peaks at relatively low
values with a large declining tail towards large periods.  The
distributions of period and SMA are flatter than for HR 8799c
because of the large range of allowed eccentricities.  The
preferred inclination is 45 $\pm$ 8$^{\circ}$.  This is fully
consistent with the values obtained for HR 8799b and c.  The
other parameters are again largely unconstrained.  The
distribution of T$_{o}$ brackets the epoch where the data was
obtained, but spans about 40 years, so does not appear to be
strongly biased as we see in the case of HR 8799e.

For HR 8799e (Figure \ref{fig:1d_e}), running the simulations
unweighted yields an eccentricity distribution skewed towards
high eccentricty, with a peak near 0.6 and an upper limit of
0.8.  Weighting significantly reduces the range of allowed
eccentricities, with an upper limit of $<$0.4.  The preference
for eccentricity between 0.1-0.2 may possibly due
to the larger astrometric uncertainties rather than an actual
elevated eccentricity.  Fits in that
region tend to favor a face-on configuration, though the
distribution extends to 40$^{\circ}$ at lower probability.  Lower and
higher eccentricities tend toward larger inclination, between
20-50$^{\circ}$.  The inclination predicted for HR
8799e in the unweighted simulations is 47 $\pm$ 6$^{\circ}$.  As
predicted, the unweighted simulations have a peak T$_{o}$ of
2015, implying that the best-fit solutions have periastron
passage happening immediately after the last data point was
taken.  The spread in the peak ranges from 2010-2020, thus
fully overlapping our data set.  Weighting instead shifts the
distribution of T$_{o}$ to the mid-1990s, with a peak in
1997.  Given the peak in the period distribution for the
weighted simulations of $\sim$35 years and an SMA peak of
$\sim$13 AU, this range of T$_{o}$ seems fairly reasonable.
In the unweighted simulations, the periods and SMAs are fairly
flat, extending out to $\sim$400 years and $\sim$100 AU,
respectively.  Weighting also moves $\Omega$ and $\omega$ from
fairly peaked at certain values to relatively unconstrained.

To assess the dependence of orbital plane configuration on
eccentricity, in Figure \ref{fig:2d_contour} we show 1- and 2-$\sigma$
contours for the joint probability density functions (PDFs) between
eccentricity and inclination and the 1-$\sigma$ contours for
eccentricity and $\Omega$ (using the weighted fits to HR 8799e).
These figures demonstrate that the orbital planes of all four planets
are consistent within $<$2$\sigma$.  They also show the
preferred values of inclination and $\Omega$ for low
eccentricity orbits.  For $\Omega$, there is overlap between
$\sim$50-70$^{\circ}$ at low eccentricity.  For inclination,
there is overlap at $\sim$30$^{\circ}$ for low eccentricity
solutions of HR 8799b, c, and d at 1-$\sigma$, and HR 8799e at 2-$\sigma$. 

Although we have not performed any dynamical analyses to
assess the possible stability of these orbital solutions, a
very rough proxy for viable solutions is to
consider only those solutions that do not come within a Hill radius of
the other planets, as shown in the left panel of Figure
\ref{fig:orb_fam}.  This analysis can be performed using our
original simulations (using the weighted version for HR
8799e), which consider each planet independently 
when fitting to assess the allowed parameter distributions
from astrometry, and the nominally preferred masses of the
planets (5, 7, 7, and 7 M$_{Jup}$ for HR 8799b, c, d, and e,
respectively).  Additional constraints on the masses of the 
planets from dynamical arguments requires simultaneous orbit modeling
that beyond the scope of this paper. Nonetheless, we wish to
get a rough sense of the distribution of parameters from our
simulations that are more likely to be stable solutions.  The distributions of orbital parameters for only
these solutions are shown in Figure \ref{fig:nocross_1d}.  Requiring orbits not
to ``cross'' significantly limits the eccentricity of the inner
planets, leading to upper limits of $<$0.3 for all four
planets.  This naturally leads to lower values for period and
SMA.  Additionally, the distribution of orbital 
inclination fully overlaps for all four planets, with values
between 30-40$^{\circ}$ consistent to within 1$\sigma$.  The
preferred inclination for HR 8799d and e moves to slightly
lower values more consistent with the distributions for HR
8799b and c.  There also seems to be a general preference for
$\Omega$ between 40-70$^{\circ}$, but larger values are still
allowed.

\section{Discussion}\label{discussion}

\subsection{Comparison to Previous Analyses}

We have computed distributions of allowed orbital parameters
for all of the HR 8799 planets using a data set for which we
have attempted to minimize systematic uncertainties in
astrometry by using the same camera and reduction
techniques.  The small reduced $\chi^{2}$ and low fit
residuals suggest that our error bars at least properly
capture remaining systematic uncertainties.

For the most part, the orbital parameters we derive are
consistent with previous fits to within the uncertainties.  We
do find a slight difference between our preferred inclinations
and those in previous works, with ours favoring values closer
to $\sim$30-40${^{\circ}}$ than to 10-30$^{\circ}$ (e.g.,
\citealt{currie12, esposito13, pueyo15}).
Still, for
all four planets, inclinations of $\lesssim$30$^{\circ}$ are
allowed, consistent with a number of previous works that
assumed a value of 28$^{\circ}$ for the system inclination.
It is notable, however, that a few analyses unrelated to orbit fitting 
predict disparate inclinations for this system.  Results from
asteroseismology suggest that an inclination of
$>$40$^{\circ}$ is preferred \citep{wright11} for the star HR
8799 itself.  Meanwhile analysis of the far infrared emissions
from the debris
disks in the HR 8799 system suggest an inclination
of $<$25$^{\circ}$ \citep{su09}, with a recent analysis from
\citet{matthews14} suggesting a disk inclination of 26 $\pm$
3$^{\circ}$.  Most recently, ALMA observations resolved the HR8799
planetessimal belt, and found an inclination of
40$^{+5}_{-6}$$^{\circ}$ \citep{booth16}.  Our analysis gives an
inclination more consistent with the asteroseismology and the ALMA results than
previous work, offering the interesting possibility that the
star, disk, and planets are in fact co-aligned.  However, our
analysis is also consistent within 1$\sigma$ with
the results from far-infrared data.  Given the large range of possible
inclinations in this and in
previous works, we believe it is premature to make claims about possible
mutual inclinations between the star, planets, and debris
disks.  It is comforting, however, that all analyses suggest
non-face on configurations for the system.       

While no statistically
significant mutual inclinations between any of the planets is
found, especially when orbits that do not cross each other's
Hill radii are considered, we are interested
in assessing the resulting orbital parameters if the planets
were forced to be coplanar.  This is similar to previous works
in which coplanarity was assumed based on dynamical arguments
(e.g., \citealt{soummer11, goz14}).  We opted to use values of inclination and $\Omega$
sampled from the solutions for HR 8799b shown in Figure
\ref{fig:1d_b}.  From these distributions, we randomly
selected 10000 values for inclination and $\Omega$ and then performed a similar Monte Carlo
simulation to those described above for HR 8799c, d, and e,
fixing those values (weighting the fits for HR 8799e).  The results of these simulations are
shown in Figure \ref{fig:mutinc_1d}.  All fits return
reasonable values of reduced $\chi^{2}$, with the best-fit
reduced $\chi^{2}$ of 0.32, 0.58, and 0.46 for HR 8799c, d, and e, respectively.  These
values are very close to our overall best fit $\chi^{2}$. 
The solutions also tend to yield relatively low eccentricities,
with the upper limit for eccentricity dropping to 0.3 for HR
8799c and d, and 0.5 for HR 8799e.  This is another indication
that coplanar solutions for \textit{all} the planets in HR
8799 are fully consistent with current astrometry.  Given the
distributions of allowed inclinations in our various orbit
fits, we believe that no offset is detectable with current
astrometry and phase coverage if it is $\lesssim$20$^{o}$ -
offsets larger than this would be detectable.

\begin{figure*}
\epsscale{0.8}
\plotone{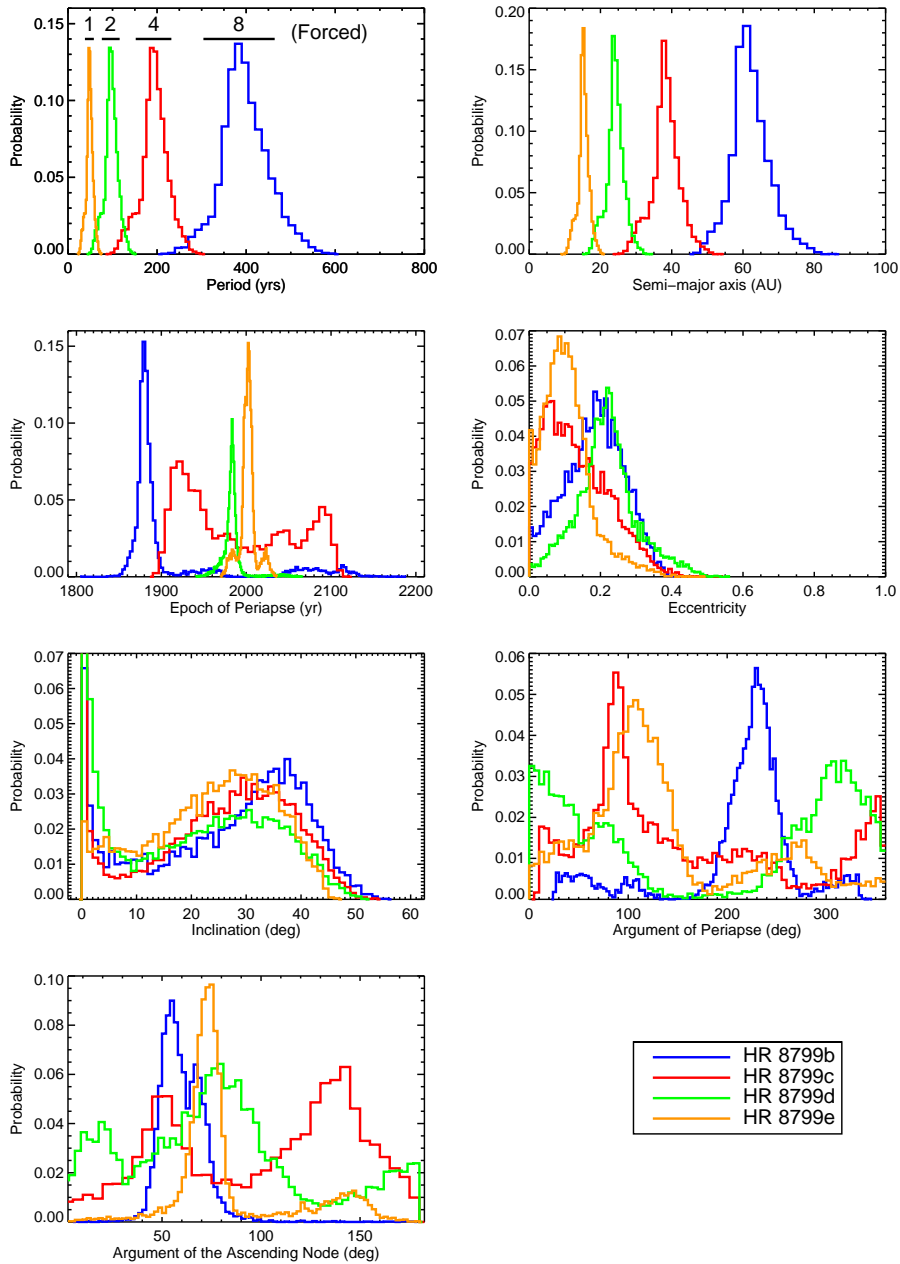}
\caption{One-dimensional PDFs for all four planets under the
  assumption that the periods are in a 1:2:4:8 resonance.  The
periods are randomly sampled from 10000 solutions for HR 8799e
in the case where the orbits are non-crossing
(Figure \ref{fig:nocross_1d}).  The periods are then multipled by the
appropriate factor for each planet.    There is also
generally a fairly fixed set of orbital parameters that fit
this criteria for HR 8799b, with strong weighting towards
specific values for T$_{o}$ and $\omega$.  In addition, resonance
is forced, higher eccentricity solutions are preferred for HR
8799b and HR 8799d than we find in our other simulations.}
\label{fig:res_1d}
\end{figure*}

Previous dynamical analysis has also suggested that the
planets may orbit in either a 1:2:4:8 resonance configuration
(e.g., \citealt{goz14}) or a 1:2:4 resonance between the inner
planets with HR 8799b not participating in any resonance
(e.g., \citealt{fabrycky10,marois10a}).  In Figures
\ref{fig:nocross_1d} and \ref{fig:mutinc_1d} we note the
locations of 1:2:4:8 period ratios assuming that the
peak and 1$\sigma$ values of the period distribution of HR 8799e is the starting
point of the chain. In both cases, the distributions for
all planets are consistent with these ratios within 1$\sigma$.  In the case of
the constraints from non-crossing Hill radii, the peak of the distribution for
HR 8799b is only marginally consistent with this resonance
chain.  Still, both sets of
analysis return results consistent with dynamical modeling predictions.

In order to further assess the orbital parameters required for the planets,
particularly HR 8799b, to be 
consistent with a 1:2:4:8 resonance in the case of
non-crossing orbits, we performed another simulation where we
used the distribution of periods from HR 8799e shown in Figure
\ref{fig:nocross_1d}, fixed a period with the appropriate
multiplicative factor for the other three planets, and then
determined the preferred orbital solution.  This analysis is
similar to that described above where the plane of the orbit
was fixed for three of the planets.  The results of this
simulation are shown in Figure \ref{fig:res_1d}.  In order to get
fits that satisfy this resonance, solutions with very specific
values of T$_{o}$ and $\omega$ are required that lead to
higher eccentricity values than in our other simulations.
However, the orbital plane parameters remain consistent with
previous simulations, and low eccentricity solutions still
allowed.  There is also a tendency for the solutions for HR
8799d to be slightly higher eccentricity than in the case of
the previous non-crossing solutions, but again lower
eccentricity solutions are allowed.  Future dynamical stability
simulations that further consider the case where HR 8799b does
not participate in a resonance with the inner three planets
could potentially yield interesting additional constraints on
its orbital configuration.

\subsection{Impact of Biases on Orbital Parameters}

For undersampled orbits such as these and for other directly
imaged exoplanets, it is important to assess whether
systematics in astrometric data sets play a roll in the output of
orbit fitting, regardless of the method of the fitting.  For
instance, poor $\chi^{2}$ values when minimal data points
exist, sampling less than 5\% of an orbital period, suggests
that data points and error bars may need to be reassessed
before conclusions are drawn about the orbital parameters.
For instance, in this work we have shown that using a dataset
in which systematics are controlled leads to slight
differences in the predicted orbital elements compared to
previous work.  We find that the inclination of HR 8799d is
not inconsistent with the other planets, in contrast to
previous work using all available astrometry
(\citealt{currie12,pueyo15}).  In order to determine whether
systematics could generate an apparent shift in the
distribution of orbital inclinations, we performed a short set
of simulations using an assumed orbit for HR 8799d.  This
orbit had an eccentricity of 0.02, an inclination of
29$^{\circ}$, and $\Omega$ of 59$^{\circ}$.  We generated a simulated data set based
on this orbit sampled at the same times as all previous
astrometric measurements for HR 8799d.  We
then estimated the possible size of systematic offsets between
data sets based on the offsets of data points from orbit fits
in \citet{soummer11}, \citet{esposito13}, \citet{pueyo15}, and
\citet{zurlo15}.  The apparent size of these offsets range
from 5 - 30 mas.  We then  applied offsets randomly sampled
from this size range to different datapoints in our
simulated astrometric data.  Our goal was not to encompass all
possible systematics, but rather to construct
a notional representation of a possible data set and see if we
could generate a significant inclination offset from the true
inclination.  We find that applying
systematics of this magnitude to the data can result in the
most likely inclination of the planet being higher than the
true value by 5-10$^{\circ}$.  However, the distribution
always encompasses the correct value of 29$^{\circ}$ such that
it would be allowed to 1-2$\sigma$.  This is somewhat
analogous to the results from \citet{pueyo15}, where there is
overlap in the inclination distributions of all four planets
at slightly greater than 1$\sigma$, but the maximum likelihood
values are off by $\sim$15$^{\circ}$.  Thus, it is important to
account for possible systematics such as those introduced from
using astrometry from multiple cameras and data pipelines in
the case where the orbits sample $\lesssim$10$\%$ of
the total period.      

\section{Summary}

We have presented both new and updated astrometric
measurements from Keck for the planets around HR 8799.  In
order to minimize systematics, we have performed orbit fits to
astrometry from Keck and NIRC2 only, and shown that the
orbital planes of the planets are consistent based on current
data.  We have also shown that the eccentricities are likely
low, and that at least the inner three planets have period
distributions consistent with a 1:2:4 resonance
configuration.  It is important to interpret the results of
orbital fits to long period systems with minimal phase coverage
with some caution, as systematics can bias the resulting
parameter distributions.

Future work will include an update to dynamical models for the
system using our improved astrometric measurements, and
continued monitoring of the system to promote additional
acceleration detections.  The data from newly
commissioned instruments like the Gemini Planet Imager and
SPHERE have been shown to yield improved astrometric error bars compared to
earlier work (e.g., \citealt{derosa15, zurlo15}), and if
properly calibrated for systematics, have 
the potential to yield stronger constraints on the orbital
parameters of this fascinating multiplanet system.  Our
continued monitoring of the system with Keck and NIRC2, with
their exquisite astrometric calibration, will play a vital
role in constraining biases in future observations with other facilities.   

\acknowledgements

The authors thank observing assistants Joel Aycock, Heather
Hershley, Carolyn Jordan, Gary Puniwai, Julie Rivera, Terry
Stickel, and Cynthia Wilburn, and support astronomers Randy
Campbell, Al Conrad, Scott Dahm, Grant Hill, Mark Kassis, Jim
Lyke, Luca Rizzi, and Hien Tran, for
their help in obtaining the observations.  We thank Eric
Nielsen, Tuan Do, and Jessica Lu for helpful conversations
about fitting methodology.  We also thank
Sylvana Yelda for helpful suggestions (7439).  And thank you
to an anonymous referee whose comments 
this manuscript.  Portions of this work were
performed under the auspices of the U.S. Department of Energy
by Lawrence Livermore National Laboratory under Contract
DE-AC52-07NA27344.  This research was supported by a NASA
Origins of Solar Systems grant to LLNL.  Portions of this work
were also performed 
at the Dunlap Institute for Astronomy \& Astrophysics,
University of Toronto. The 
Dunlap Institute is funded through an endowment established by
the David Dunlap family and the University of Toronto.  The
W.M. Keck Observatory is operated as 
a scientific partnership among the California Institute of
Technology, the University of California and the National
Aeronautics and Space Administration. The Observatory was made
possible by the generous financial support of the W.M. Keck
Foundation.  The authors also wish to recognize and
acknowledge the very significant cultural role and reverence
that the summit of Maunakea has always had within the
indigenous Hawaiian community.  We are most fortunate to have
the opportunity to conduct observations from this mountain.

\clearpage
\begin{turnpage}
\begin{deluxetable}{lccccccccccc} 
\tabletypesize{\scriptsize} 
\tablecolumns{12}
\tablewidth{0pc} 
\tablecaption{HR 8799 Relative Astrometry} 
\tablehead{ 
  \colhead{} & \multicolumn{2}{c}{HR 8799b} & \colhead{} &
  \multicolumn{2}{c}{HR 8799c} & \colhead{} &
  \multicolumn{2}{c}{HR 8799d} & \colhead{} & \multicolumn{2}{c}{HR 8799e} \\
\cline{2-3}
\cline{5-6}
\cline{8-9}
\cline{11-12}\\
\colhead{Date (UT)} & \colhead{$\Delta$x (as)} &
\colhead{$\Delta$y (as)} & \colhead{} &
  \colhead{$\Delta$x (as)} & \colhead{$\Delta$y (as)} & \colhead{} &
  \colhead{$\Delta$x (as)} & \colhead{$\Delta$y (as)} & \colhead{} &
  \colhead{$\Delta$x (as)} & \colhead{$\Delta$y (as)} 
 }
\startdata 
2004 Jul 14\tablenotemark{a} & -1.471 $\pm$ 0.006 & 0.884 $\pm$ 0.006 & & 0.739 $\pm$ 0.006 & 0.612 $\pm$ 0.006 & & n/a & n/a & & n/a & n/a \\
2007 Aug 02 & -1.504 $\pm$ 0.003 & 0.837 $\pm$ 0.003 & & 0.683 $\pm$ 0.004 & 0.671 $\pm$ 0.004 & & 0.179 $\pm$ 0.005\tablenotemark{a} & -0.588 $\pm$ 0.005\tablenotemark{a} & & n/a & n/a \\
2007 Oct 25 & -1.500 $\pm$ 0.007 & 0.836 $\pm$ 0.007 & & 0.678 $\pm$ 0.007 & 0.676 $\pm$ 0.007 & & 0.175 $\pm$ 0.010 & -0.589 $\pm$ 0.010 & & n/a & n/a \\
2008 Sep 18 & -1.516 $\pm$ 0.004 & 0.818 $\pm$ 0.004 & & 0.663 $\pm$ 0.003 & 0.693 $\pm$ 0.003 & & 0.202 $\pm$ 0.004 & -0.588 $\pm$ 0.004 & & n/a & n/a \\
2009 Jul 30 & -1.526 $\pm$ 0.004 & 0.797 $\pm$ 0.004 & & 0.639 $\pm$ 0.004 & 0.712 $\pm$ 0.004 & & 0.237 $\pm$ 0.003 & -0.577 $\pm$ 0.003 & & 0.306 $\pm$ 0.007 & -0.211 $\pm$ 0.007 \\
2009 Aug 01\tablenotemark{b} & -1.531 $\pm$ 0.007 & 0.794 $\pm$ 0.007 & & 0.635 $\pm$ 0.009 & 0.722 $\pm$ 0.009 & & 0.250 $\pm$ 0.007 & -0.570 $\pm$ 0.007 & & 0.318 $\pm$ 0.010 & -0.195 $\pm$ 0.010 \\
2009 Nov 01 & -1.524 $\pm$ 0.010 & 0.795 $\pm$ 0.010 & & 0.636 $\pm$ 0.009 & 0.720 $\pm$ 0.009 & & 0.251 $\pm$ 0.007 & -0.573 $\pm$ 0.007 & & 0.310 $\pm$ 0.009 & -0.187 $\pm$ 0.009 \\
2010 Jul 13 & -1.532 $\pm$ 0.005 & 0.783 $\pm$ 0.005 & & 0.619 $\pm$ 0.004 & 0.728 $\pm$ 0.004 & & 0.265 $\pm$ 0.004 & -0.576 $\pm$ 0.004 & & 0.323 $\pm$ 0.006 & -0.166 $\pm$ 0.006 \\
2010 Oct 30 & -1.535 $\pm$ 0.015 & 0.766 $\pm$ 0.015 & & 0.607 $\pm$ 0.012 & 0.744 $\pm$ 0.012 & & 0.296 $\pm$ 0.013 & -0.561 $\pm$ 0.013 & & 0.341 $\pm$ 0.016 & -0.143 $\pm$ 0.016 \\
2011 Jul 21 & -1.541 $\pm$ 0.005 & 0.762 $\pm$ 0.005 & & 0.595 $\pm$ 0.004 & 0.747 $\pm$ 0.004 & & 0.303 $\pm$ 0.005 & -0.562 $\pm$ 0.005 & & 0.352 $\pm$ 0.008 & -0.130 $\pm$ 0.008 \\
2012 Jul 22 & -1.545 $\pm$ 0.005 & 0.747 $\pm$ 0.005 & & 0.578 $\pm$ 0.005 & 0.761 $\pm$ 0.005 & & 0.339 $\pm$ 0.005 & -0.555 $\pm$ 0.005 & & 0.373 $\pm$ 0.008 & -0.084 $\pm$ 0.008 \\
2012 Oct 26 & -1.549 $\pm$ 0.004 & 0.743 $\pm$ 0.004 & & 0.572 $\pm$ 0.003 & 0.768 $\pm$ 0.003 & & 0.346 $\pm$ 0.004 & -0.548 $\pm$ 0.004 & & 0.370 $\pm$ 0.009 & -0.076 $\pm$ 0.009 \\
2013 Oct 16 & -1.545 $\pm$ 0.022 & 0.724 $\pm$ 0.022 & & 0.542 $\pm$ 0.022 & 0.784 $\pm$ 0.022 & & 0.382 $\pm$ 0.016 & -0.522 $\pm$ 0.016 & & 0.373 $\pm$ 0.013 & -0.017 $\pm$ 0.013 \\
2014 Jul 17 & -1.560 $\pm$ 0.013 & 0.725 $\pm$ 0.013 & & 0.540 $\pm$ 0.013 & 0.799 $\pm$ 0.013 & & 0.400 $\pm$ 0.011 & -0.534 $\pm$ 0.011 & & 0.387 $\pm$ 0.011 & 0.003 $\pm$ 0.011 \\
\enddata
\tablenotetext{a}{This data set was taken in non-ADI mode and
  thus was not reprocessed with SOSIE.  We include the values
  here from \citet{marois08} for completeness, as it is NIRC2
  data that we included in our orbit fitting.  Only HR 8799b and c
  are detected in this data set.}
\tablenotetext{b}{Due to the proximity of HR 8799d to the 1000
mas focal plane mask in this epoch, we believe its position is
biased.  We therefore elected not to include it 
in orbit fitting.}
\tablenotetext{b}{This epoch was not used for orbit
  fitting due to the close time sampling to the 2009 July 30
  points.  We include the astrometry here for completeness.}
\tablecomments{HR 8799e is not detected in any data taken
  prior to 2009. In the original publication of data from 2007
  and 2008 in \citet{marois08}, there was an error in the
  application of the offset with respect to North.  This has been remedied here.}
\label{tab:astr}
\end{deluxetable}

\end{turnpage}

\end{document}